\documentclass[preprint,12pt,nonatbib]{elsarticle}

\makeatletter
\AtBeginDocument{%
  \expandafter\gdef\csname X@deceased\endcsname{%
    \ensuremath{\dagger}%
  }%
}
\newcommand{\deceasedref}{\fnref{deceased}}
\newcommand{\deceasedtext}[1]{%
  \g@addto@macro\@fnotes{%
    \begingroup
      \def\thefootnote{\textdagger}%
      \footnotetext{#1}%
    \endgroup
  }%
}
\makeatother

\usepackage[
  backend=biber,
  style=numeric,      
  sorting=none,       
  maxbibnames=3,      
  minbibnames=3       
]{biblatex}
\usepackage{amssymb}
\usepackage{xspace}

\usepackage[inkscapelatex=false]{svg}

\usepackage{lineno}
\usepackage[dvipsnames]{xcolor}
\usepackage{caption}
\usepackage{subcaption}

\title{The High-Energy Light Isotope eXperiment (HELIX) Instrument\\
} 
\author[osu]{P.~S.~Allison}
\author[quu]{M.~Baiocchi\fnref{fn1}}
\author[osu]{J.~J.~Beatty}
\author[uch]{L.~Beaufore\fnref{fn2}}
\author[osu]{D.~H.~Calder\'on}
\author[psu]{Y.~Chen\fnref{fn3}}
\author[psu]{S.~Coutu}
\author[mcg]{E.~Ellingwood\fnref{fn4}}
\author[uch]{D.~Fuehne}
\author[umi]{N.~Green}
\author[mcg]{D.~Hanna\corref{cor1}} 
\author[uch]{H.~B.~Jeon\fnref{fn5}}
\author[inu]{B.~Kunkler\deceasedref}
\author[inu]{M.~Lang\deceasedref}
\author[uch]{R.~Mbarek\fnref{fn6}} 
\author[uch]{K.~McBride}
\author[quu]{C.~E.~McGrath}
\author[psu]{I.~Mognet}
\author[inu]{J.~Musser}
\author[nku]{S.~Nutter}
\author[mcg]{S.~O'Brien\fnref{fn7}}
\author[quu]{N.~Park}
\author[mcg]{T.~Rosin\fnref{fn8}}
\author[uch]{K.~Sakai}
\author[cba]{M.~Tabata}
\author[umi]{G.~Tarl\'e}
\author[inu]{G.~Visser}
\author[uch]{S.~P.~Wakely}
\author[uch]{I.~G.~Wisher}
\author[psu]{M.~Yu\fnref{fn9}}

\affiliation[osu]{The Ohio State University, Department of Physics, Columbus, OH, USA}
\affiliation[quu]{Queen's University, Department of Physics, Engineering Physics & Astronomy, Kingston, ON, Canada}
\affiliation[uch]{University of Chicago, Enrico Fermi Institute,  Chicago, IL, USA}
\affiliation[psu]{Pennsylvania State University, Department of Physics, University Park, PA,  USA}
\affiliation[mcg]{McGill University, Department of Physics,  Montreal, QC, Canada}
\affiliation[umi]{University of Michigan, Department of Physics, Ann Arbor, MI, USA}
\affiliation[inu]{Indiana University, Department of Physics,  Bloomington, IN, USA}
\affiliation[nku]{Northern Kentucky University, Department of Physics, Geology and Engineering Technology, Highland Heights, KY, USA}
\affiliation[cba]{Chiba University, Department of Physics, Chiba, Japan}

\newpageafter{author} 
\cortext[cor1]{Corresponding author}
\fntext[fn1]{Now at Gran Sasso Science Institute, L’Aquila, Italy}
\fntext[fn2]{Now at the Ohio State University, Columbus, OH, USA}
\fntext[fn3]{Now at UCLA, Los Angeles, CA, USA}
\fntext[fn4]{Now at University of Edinburgh, Edinburgh, UK}
\fntext[fn5]{Now at NASA GSFC, Greenbelt, MD, USA}
\fntext[fn6]{Now at Princeton University, Princeton, NJ, USA}{}
\fntext[fn7]{Now at CertX Canada, Montreal, QC, Canada}
\fntext[fn8]{Now at MY01 Inc., Montreal, QC, Canada}
\fntext[fn9]{Now at Purple Mountain Observatory, Nanjing, China}
\deceasedtext{Deceased.}

\newcommand{\usim}{\mathord{\sim}}
\newcommand{\um}{$\mu$m\xspace}
\newcommand{\ie}{\textit{i.e.}}
\newcommand{\eg}{\textit{e.g.}}

\begin{document}

\begin{frontmatter}

\begin{abstract}
The HELIX detector is a balloon-borne instrument designed to measure the flux of light ($Z<11$) cosmic-ray isotopes. In this paper we describe the initial configuration of HELIX, optimized for measurements in the energy range from approximately 0.2 GeV per nucleon to beyond 3 GeV per nucleon. In addition to a spectrometer, which is based on a one-tesla superconducting magnet and a drift-chamber tracker, HELIX employs scintillator-based time-of-flight counters and a ring-imaging Cherenkov detector, allowing event-by-event reconstruction of primary particle mass, charge, and magnetic rigidity.
\end{abstract}

\end{frontmatter}

\section{Introduction}
\label{sec:intro}

The High Energy Light Isotope eXperiment (HELIX)~\cite{2019ICRC...36..121P} has been developed to measure the chemical and isotopic abundances of light (\ie, atomic number $Z<11$) cosmic-ray nuclei. 
The primary goal of the HELIX program is to measure the $^{10}$Be/$^9$Be flux ratio over the energy range from $\usim 0.2$ GeV/n to, eventually, $\usim 10$ GeV/n. 
\\

The detector incorporates a drift-chamber tracker embedded in a one-tesla superconducting magnet to measure particle rigidities. 
A system of time-of-flight scintillation detectors is used for triggering and to measure velocities at energies up to $\usim 1$~GeV/n.  A ring-imaging Cherenkov detector is used for velocity at higher energies. By combining its charge $Z$, measured with the ToF detectors, with its mass $m = \frac{RZe}{\gamma \beta c^2}$, calculated from the rigidity $R$ and velocity $\beta$, one can uniquely identify the isotopic nature of each incident particle.\\

Due to the design of the magnet, with its open geometry, incoming particles encounter very little material before their rigidities are measured. Summing over the upper scintillators, support structures, the top of a thin pressure vessel that contains the DCT, and internal tracker components yields a total grammage of approximately $2~g/cm^2$.\\

\begin{figure}[htbp]
\centering
\centering
\includegraphics[width=0.99\textwidth]{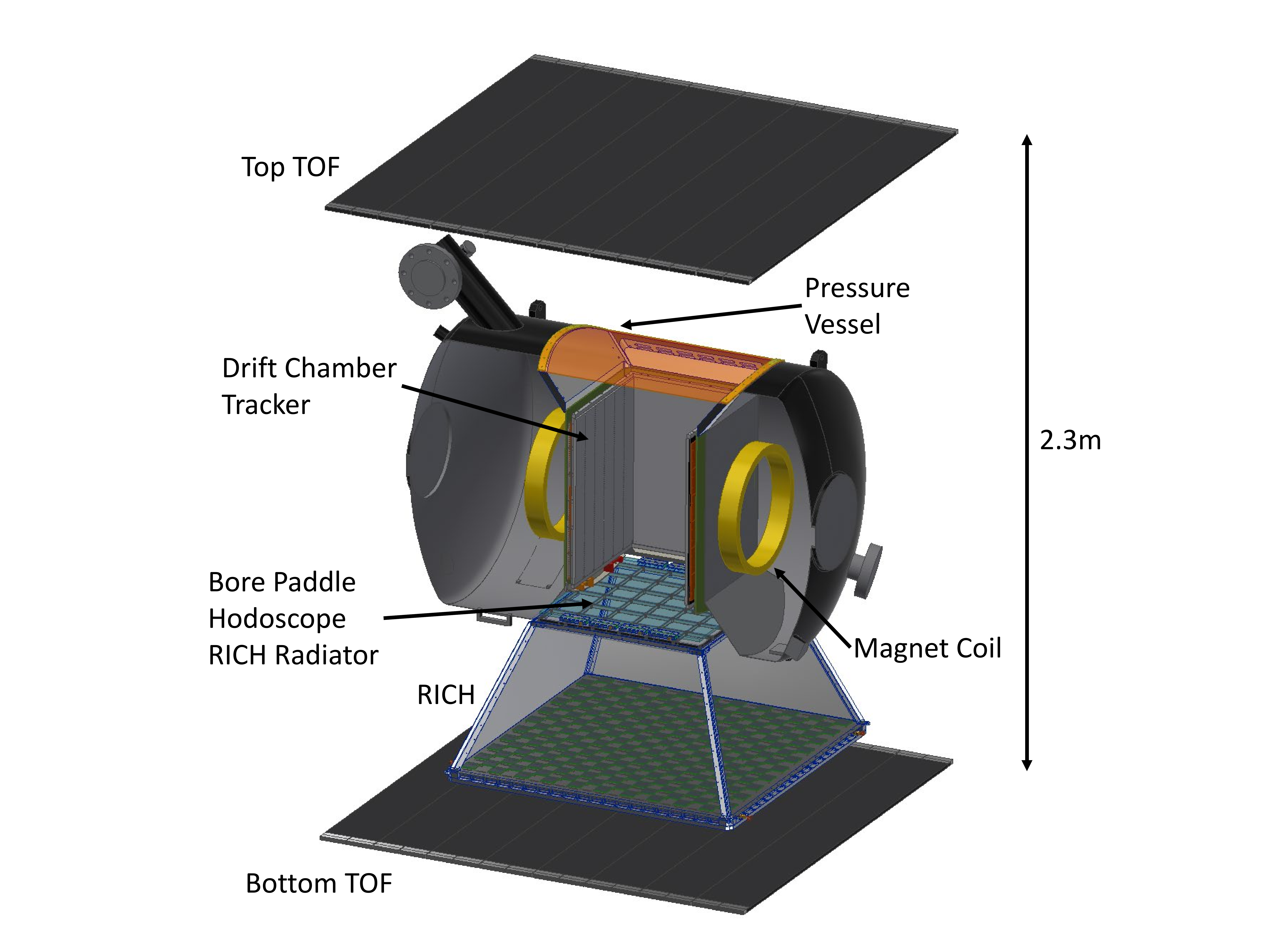}
\vskip 0cm
\caption{
Rendering of the HELIX instrument, showing the ToF counters above and below the cryostat housing the one-tesla superconducting magnet. The magnet coils, shown in yellow, bracket the DCT installed in the magnet bore. At the bottom of the bore is a plastic scintillator counter followed by a scintillating-fiber hodoscope. The RICH detector, comprising an aerogel radiator plane and an SiPM readout plane, is located immediately below. See the text for details.}


\label{instrument}
\end{figure}

A rendering of the instrument is shown in Fig.~\ref{instrument}. The various components will be described in detail in the following sections and some preliminary performance results, obtained during the NASA Columbia Scientific Balloon Facility (CSBF) 2024 arctic balloon campaign, will be shown. In that campaign, the payload successfully undertook a six-day engineering flight from Esrange, Sweden (67$^\circ$53$'$ N, 21$^\circ$05$'$ E), to Ellesmere Island in the Canadian Arctic (78$^\circ$08$'$ N, 80$^\circ$49$'$ W).\\

An overview of the design considerations for the HELIX instrument is presented in Sec.~\ref{sec:design}. Sec.~\ref{sec:magnet} discusses the superconducting magnet design, while Sec.~\ref{sec:DCT} presents an overview of the drift-chamber tracker (DCT), located in the magnet's warm bore. The time-of-flight (ToF) system  and ring imaging Cherenkov (RICH) detector are discussed in Sec.~\ref{sec:TOF} and \ref{sec:RICH} respectively, while the hodoscope, which provides additional tracking in the non-bending plane, is discussed in Sec.~\ref{sec:hodo}. Finally, Sec.~\ref{sec:Trigger} discusses the trigger and data acquisition system, and presents an overview of the power system.


\section{Design Considerations\label{sec:design}}
HELIX is designed as a conventional particle telescope, with three triggering planes to define the geometric acceptance.  The overall geometry is determined by balancing the sometimes conflicting requirements of large geometric factor for high particle flux, large ToF spacing for the desired velocity resolution, total payload mass, which affects the maximum payload altitude, and overall geometric size, which is subject to the constraints of the CSBF launch vehicle. The payload gondola is designed as a lightweight lattice frame of tubular aluminum extrusion with low-mass aluminum honeycomb panels for the primary horizontal decks. Detector metrology was performed using optical photogrammetry before launch to establish relative detector positions to better than one millimeter with further refinement possible using straight-through particle trajectories collected on the last day of the flight, while the magnet was de-energized.\\

\section{Superconducting Magnet}
\label{sec:magnet}

Central to the HELIX design is a one-tesla superconducting magnet. 
This component is key to the goal of achieving the momentum resolution necessary to differentiate between different isotopes, such as $^9$Be and $^{10}$Be, which is the primary scientific goal of the project.\\

The magnet was custom-built (by Cryomagnetics Inc.) for the High Energy Antimatter Telescope (HEAT) balloon-borne instrument~\cite{1997NIMPA.400...34B}, at which time it was operated within a large pressure vessel. The magnet was refurbished for HELIX, to operate without such a pressure vessel under stratospheric conditions, and its control, housekeeping and discharge systems were replaced with updated technology.\\

The magnet features two niobium–titanium coils, approximately 0.47~m in diameter and separated by 0.73~m, bracketing a warm-bore volume of 0.51 m $\times$ 0.51 m $\times$ 0.61 m. The design current of 91.7 A produces a 1~T central field. The combined mass of the aluminum cryostat and 260 L of $\usim4.2$ K liquid helium at launch is approximately 460~kg.  During the 2024 flight, a hold time of approximately 5.2 days was accumulated prior to discharging the magnet in order to collect a sample of magnet-off events before the flight was terminated.   

\begin{figure}[htbp]
\centering
\includegraphics[width=0.49\textwidth]{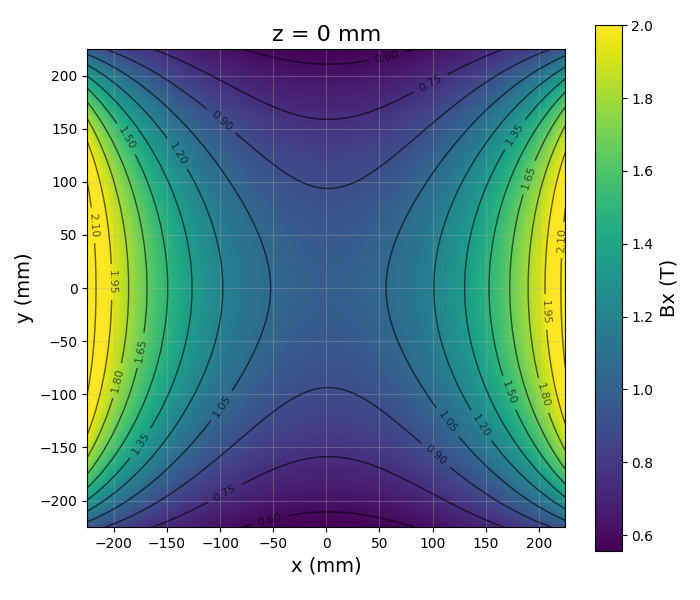}
\includegraphics[width=0.49\textwidth]{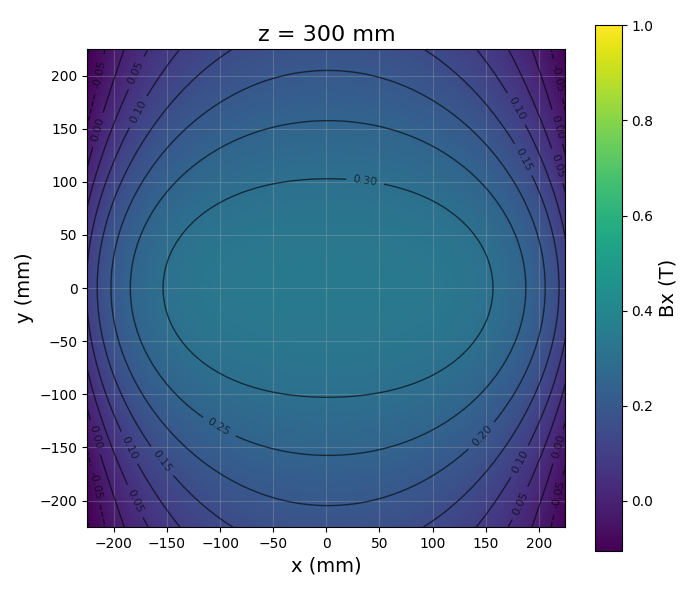}
\caption{
Maps of the x-component of the magnetic field (B$_x$) for two planes in the magnet bore.  The plot on the left shows z=0 mm (bore center) and the plot on the right shows the field at z=300 mm (top of bore). }
\label{B-field}
\end{figure}

HELIX uses a right-handed Cartesian coordinate system defined such that its x axis passes through the centers of the magnet coils along the direction of the field and x=0 is located midway between the coils. The z axis points vertically upwards. Fig.~\ref{B-field} shows maps of the magnitude of the x-component of the magnetic field in the region of the magnet bore. The path-integrated field, important for calculating the maximum detectable rigidity of a particle, is typically 0.42 T-m for tracks passing through the bore.  




\section{Drift-chamber Tracker}
\label{sec:DCT}

HELIX incorporates a drift-chamber tracker to measure the trajectories of cosmic-ray nuclei passing through the instrument.
Its primary purpose is to determine the rigidity ($ R = pc/Ze $) of each track by measuring the curvature caused by the field of the superconducting magnet.
Track information from the DCT is also important for analyzing data from the RICH. The coordinates of the track extrapolated to the readout plane are also used as an estimate of the center of the Cherenkov ring.\\

\subsection{DCT Design}

\begin{figure}
    \centering
    \includegraphics[width=0.99\linewidth]{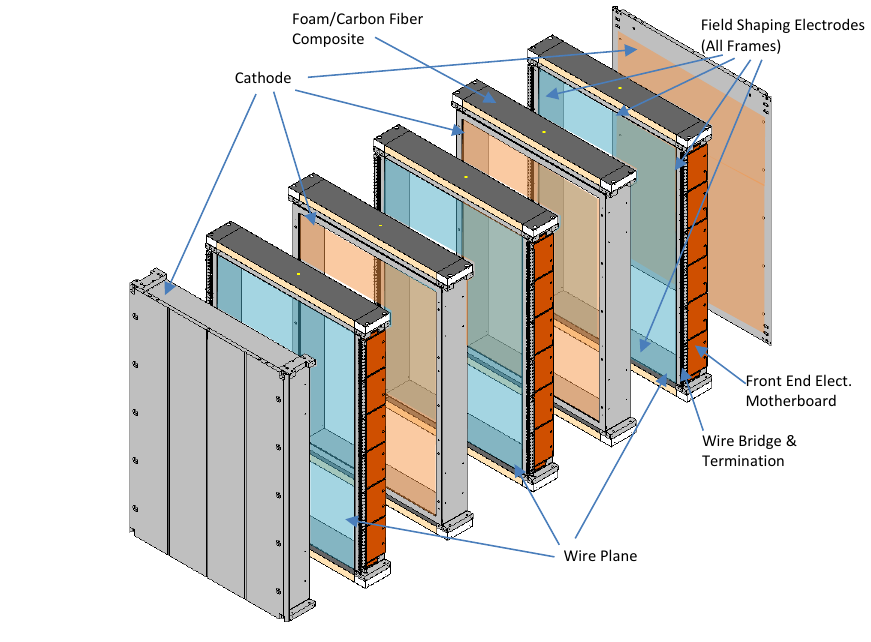}
    \caption{An exploded view of the DCT, illustrating its mechanical construction. There are three cells, each with an anode plane located between two drift regions that are bound by cathode planes. The anode planes, shown in blue, are each made from a series of 72 sense wires and 76 potential wires (see Fig.~\ref{fig:DCT-fig}) and the cathode planes, shown in tan, are either stretched copper mesh (inner two modules) or a continuous copper layer (outer two modules). The vertical sides of the modules are rectangles of G-10 material, 12.7 mm thick,  while the top and bottom sides are made from a low-mass foam/carbon-fiber composite. }
    \label{fig:DCT-explode}
\end{figure}

The DCT extends 450 mm $\times$ 450 mm laterally, is 580 mm high, and is installed in an aluminum (Al 6061-T6 alloy) pressure vessel (Meyer Tool and Manufacturing, Inc.) that fits into the magnet bore.
The vessel, rated to 2.5 atmospheres absolute, allows operation of the DCT at a pressure of 1.0 atmosphere while at float altitude, where the ambient pressure is only a few mbar. The top of the pressure vessel (1/8-inch Aluminum 6061) incorporates 48 signal feedthroughs, two gas-line ports and two high-voltage bulkhead connectors. Half of the signal feedthroughs are on the y $= 225$~mm  edge of the vessel and the other half are on the y $= -225$~mm edge. \\

\begin{figure}
    \centering
    \includegraphics[width=0.99\linewidth]{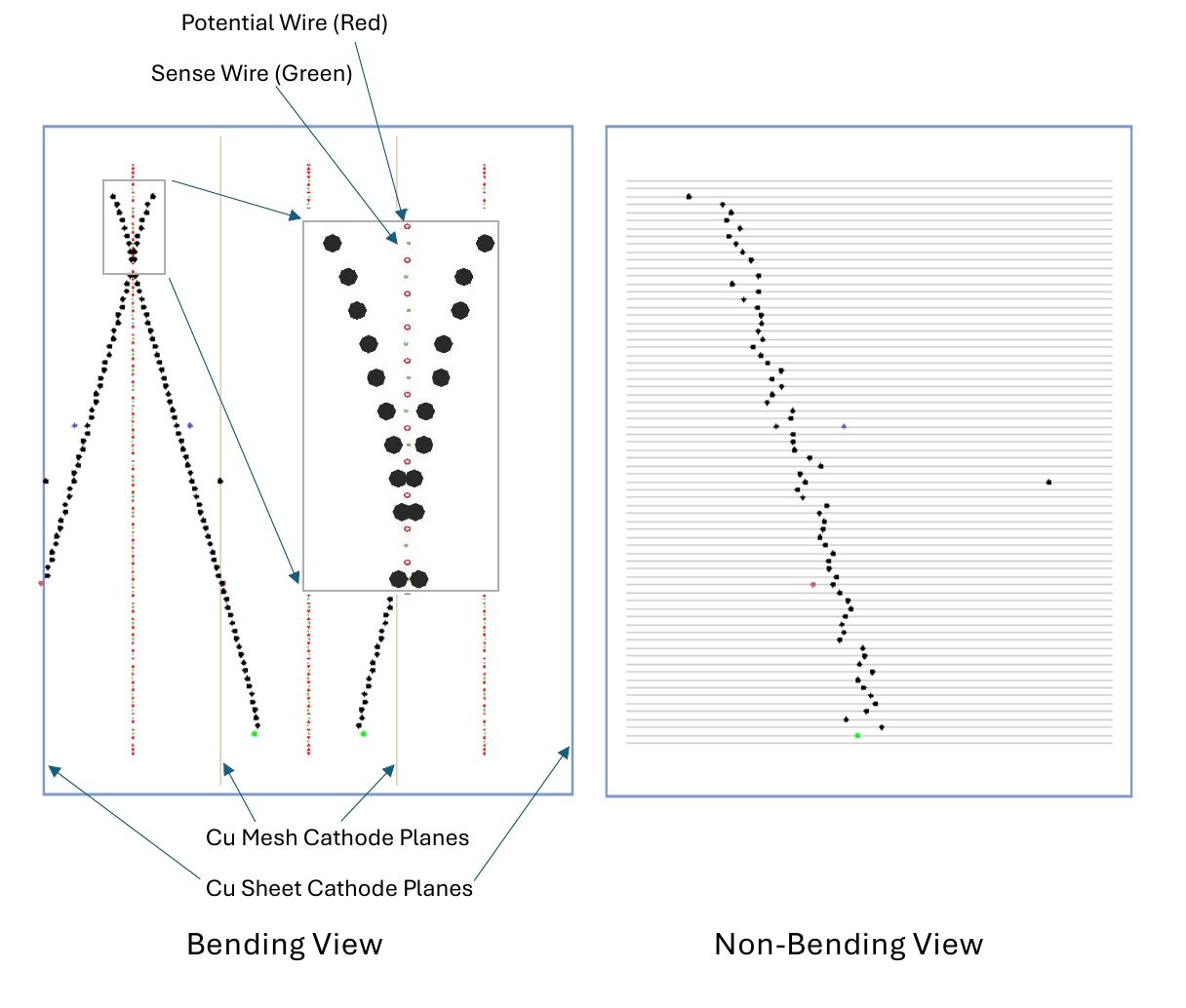}
    \caption{Two views of the drift-chamber tracker operation. The event shown is obtained from flight data, and is typical of the DCT data quality obtained during flight. The left panel shows the bending-plane view, perpendicular to the direction of the magnetic field. The dots represent points, reconstructed from drift times and wire coordinates, for a typical track. (The dots have a coarse color-code to display pulse size: black for large pulses, red for medium and green for small.) The left-right ambiguity arising in trackers of this kind gives rise to the two sets of points but can be resolved using the fact that the sense wires are staggered by $\pm$ 0.3 mm, as shown in the inset.
    The right panel shows the coordinates of the same track as seen in the non-bending plane, parallel to the magnetic field direction. The points are determined by comparing pulse sizes at opposite ends of the resistive sense wires.}

    \label{fig:DCT-fig}
\end{figure}

\subsection{Electrostatics}

The tracker, shown schematically in Fig.~\ref{fig:DCT-explode}, has a jet-chamber layout~\cite{1982NIMPR.196..293H, Opal_JET} with wires parallel to the magnetic field direction. They are configured in columns, defining three cells, as shown on the left view of Fig.~\ref{fig:DCT-fig}.
The cells are each 150~mm wide and 580~mm high, with an anode plane made from 72 sense wires at the center. The sense wires use Stablohm 800 non-magnetic resistive alloy (California Fine Wire Co.), a non-magnetic nickel-chromium alloy with a resistivity of 1.33~$\mu \Omega \cdot$m.  The sense wires are 20~$\mu m$ in diameter and are positioned with an 8~mm pitch.  They are alternately staggered by 0.3~mm from a central plane to aid in resolving the left-right ambiguity inherent in jet-chamber design. A gold-plated aluminum potential wire, 250~$\mu m$ in diameter, is located between each pair of sense wires, with an extra set of two (three) installed at the top (bottom) of each column to reduce drift-field distortions at the chamber edges.\\



There are four cathode planes for the three drift cells.
The two that define the left and right sides of the DCT volume are continuous surfaces made from a thin layer of copper foil, while the cathode planes dividing the cells from each other are made from copper mesh with a thread diameter of 2~mil ($\sim50$ \um) on 3~mil ($\sim75$ \um) centers. \\

\begin{figure}
    \centering
    \includegraphics[width=0.99\linewidth]{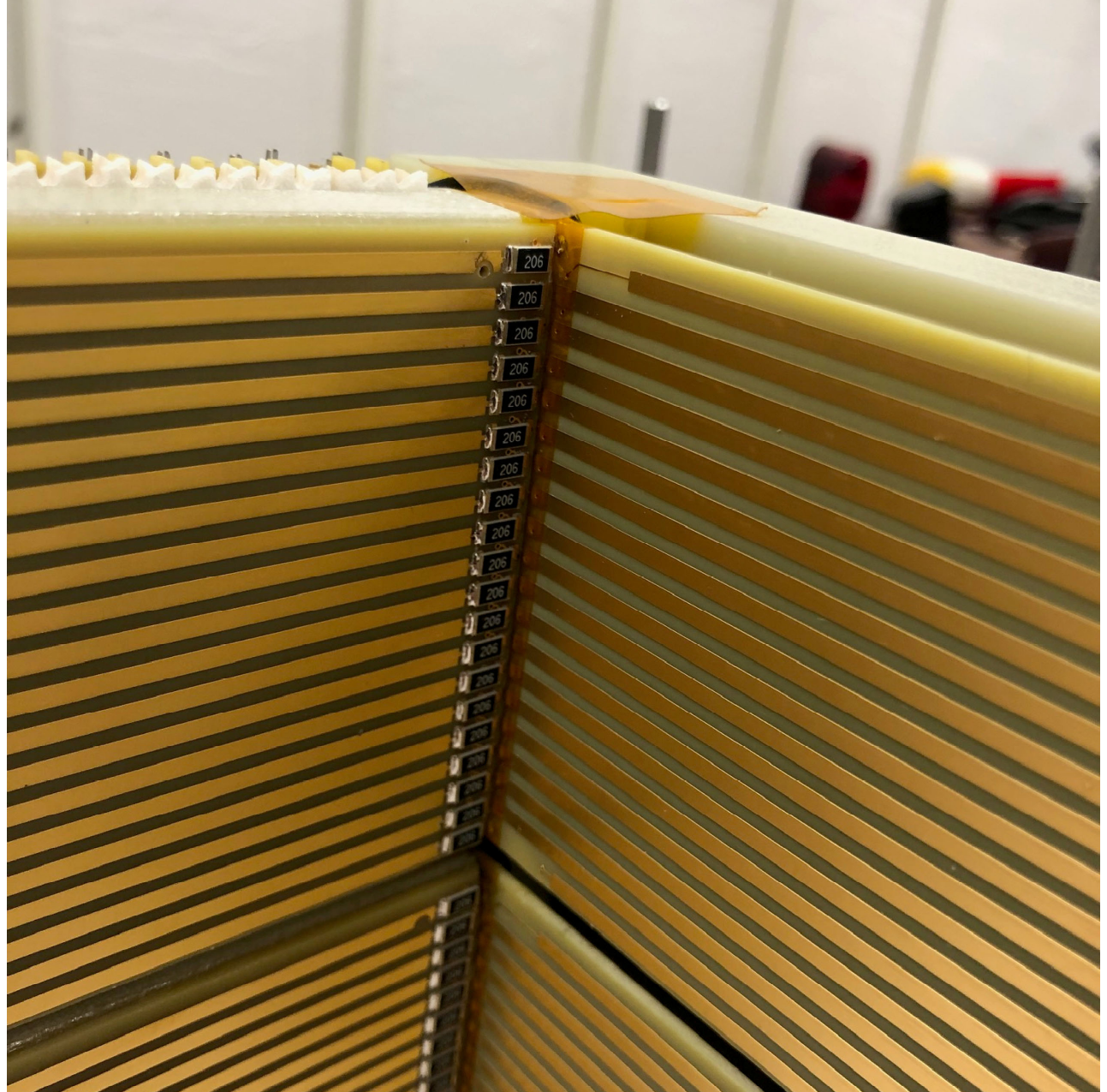}
    \caption{Interior view of the DCT, showing the field-shaping electrodes. 
    Strips on the facing surface of the printed circuit board align with gaps on the back side of the board and vice versa.
The objects positioned in a vertical line on the left side of the corner are resistors that form the voltage divider that defines the drift-field gradient.
The wire positioning bridge, formed from two precision-machined Macor plates, can be seen at the top of the figure. }
    \label{fig:DCTinnerds}
\end{figure}


The four surfaces that surround the drift volume between wire and cathode planes are instrumented with electrodes that create a uniform drift field. These electrodes are gold-plated copper strips, 2.2~mm wide, alternating on the two sides of a G-10 fiberglass epoxy laminate sheet, 1.6~mm thick.  The 42 electrode strips between each sense wire/cathode plane pair are connected to nodes of a voltage divider operating between ground and cathode plane potential, providing drift-field uniformity and proper field termination at the drift-volume boundaries.  Fig.~\ref{fig:DCTinnerds} is a photograph of the interior of the DCT taken prior to wire installation, showing the field-shaping electrodes and the resistive divider that defines the field gradient. The Macor wire positioning bridge can also be seen at the top of this figure. \\

\subsection{Mechanics}

Each sense or cathode plane is built as a separate self-supporting mechanical structure, as shown in Fig.~\ref{fig:DCT-explode}. The structures are made from two G-10 side plates, each 75 mm wide, 580 mm high and 12.7 mm thick, and a pair of low-mass foam/carbon-fiber panels forming the top and bottom sides. The G-10 sides hold the wire terminations and front-end electronics, or the edges of cathode meshes. The outermost cathode planes are bonded to 12.7~mm thick G-10 plates.  The sense-wire and cathode-plane assemblies, along with the outer cathode plates, are precision pinned to each other and tied together to form a single mechanical assembly, using four tie-rods in compression at the corners of the assembly.  This design architecture allows simple and complete access to any given sense wire or cathode plane.    \\

The primary mechanical termination of sense and potential wires is provided by feedthroughs located on the wire mounting boards, using a low outgassing UV-cured epoxy (Masterbond UV22DC80-1).  Electrical connection to the sense and potential wires is achieved through a solder connection to posts on the wire termination boards. Wires are precisely positioned by V-shaped notches formed from two glass ceramic (Macor) plates, machined to 50 $\mu$m precision.  The sense wires are strung to a tension of 20 g, the potential wires to 102 g.\\ 

The mass of the DCT pressure vessel is approximately 50 kg, while the DCT itself has a similar mass of about 50 kg.
Vertical tracks that do not pass through any wires encounter a total material thickness of less than 
$0.1~\mathrm{g/cm}^2$.


\subsection{DCT Electronics}

Both ends of each sense wire are read out, to obtain hit positions along the sense wires using the charge-division method.  The wires are terminated on front-end electronics boards located on the DCT structure, where filtering and amplification of signals, 12 channels per board, occurs. Each channel comprises a trans-impedance stage (Analog Devices LTC6268) with a gain of 20 mV/$\mu$A, followed by a pulse-shaping network and a differential output driver stage (Texas Instruments THS4521). The filter network consists of an RC low-pass filter stage with a time constant of 20 ns, followed by a second low-pass filter with a time constant of 13.3 ns, optimized for leading-edge timing resolution. The output differential-pair signals exit the DCT pressure vessel via vacuum-rated feedthroughs, and are digitized on analog-to-digital (ADC) boards located in an electronics crate near the vessel. Each of the 36~signal cables running from the DCT vessel to the ADC electronics provides 12~differential-pair sense-wire signals, power for the front-end electronics, and a test-pulse input from the ADC board. Separate cables provide readout of 24~thermistors situated around the DCT frame and gas volume, and DC power for 20 thin kapton heaters mounted on the outer frame surfaces. The heaters are used to minimize temperature gradients in the gas volume produced by the internal electronics and external thermal environment.\\

The DCT ADC system consists of 9 boards, each with three low-power 16-channel 12-bit flash ADCs (FADCs, Texas Instruments ADS52J90) running at 80 mega-samples per second (MSPS). The FADCs are configured and read out using an Artix-7 field-programmable gate array (FPGA, one per board) which also handles synchronous triggering (see Sec. \ref{sec:Trigger}).\\

The ADC boards are cabled such that both ends of each wire are read out by the same board, which enables smart zero suppression at the FPGA level, based on summed signal levels, restricting readout to regions of interest (RoIs).  The RoI signal threshold, pre-sampling length, and length of the RoI are all parameters that can be adjusted during flight as needed. Under normal operating conditions, the RoI consists of the 6 waveform samples immediately prior to the threshold crossing, and the 24 samples following the threshold crossing, for a total of 375 ns of waveform acquisition for each signal above threshold. The ADC boards also provide the ability to generate test pulses to the front-end electronics, used for functional testing.\\

\subsection{DCT Support Subsystems}


The DCT uses a 90:10 CO$_2$/Ar mixture, which is produced using an on-board gas-mixer system supplied by 
two separate lightweight aluminum 6061 cylinders,
each with 2350 L capacity. This gas mixture was chosen for its low electron-diffusion rates.  Gas flow is through high-purity polytetrafluoroethylene (PTFE)  lines connecting the gas bottles to the gas panel inlets, and the gas panel outlets to the DCT vessel, or stainless steel tubing on the gas panel interconnects between gas mixer modules and gauges.  The relative CO$_2$/Ar gas flow rates and DCT vessel pressure are controlled using MCW-series Alicat Whisper flow controllers. A precision sapphire orifice at the vessel gas outlet controls and limits the overall flow rate. At the flow rates used (typically 100 ccm), the system can supply the DCT for approximately 16 days. During flight the flow controllers were tuned to maintain a constant 90:10 CO$_2$/Ar ratio at a pressure of 14.6~psia. \\



 Operation of the DCT requires two negative high-voltage sources, one for the cathode, and the second for the potential wires.  The anodes (sense wires) are held at ground potential through 10 M$\Omega$ current-limiting resistors.  The cathodes are held at $-7500$~V, establishing a nominal drift field of 730~V/cm, as determined by simulations of the gas and DCT wire geometry with the software package \texttt{Garfield} \cite{Veenhof:1998,GarfieldPP}.   The potential wires are held at $-2700$~V, which provides a gas gain sufficient to provide high tracking efficiency for singly-charged minimum-ionizing particles. The high-voltage system is based around a pair of identical Spellman UM8-40 high-voltage supplies housed inside a small custom-built pressure vessel held at atmospheric pressure (1 atm).  The high-voltage cables connecting the high-voltage module to the DCT vessel are run though a vacuum-tight conduit with gas-tight connections at the HV module and DCT vessel, maintaining these cables at atmospheric pressure as well.  Both supplies include additional low-pass filtering of the high-voltage outputs to the DCT, designed to achieve a ripple of less than 0.5 V on both supplies at full operating voltages.\\

To provide active temperature control of the DCT volume, twelve 2.5~W and eight 4~W polyimide film heaters (Omega Engineering KHA series) are attached to the external DCT faces.  The bottom and top faces each have four 4~W heaters and the `East' and `West'\footnote{`East' and `West' here do not refer to geographical directions, but rather are designated in reference to the `North' and `South' poles of the superconducting magnet.} faces each have six 2.5~W heaters.  The total heating power is $\sim60$~W.  Twenty-four precision thermistors (Omega Engineering PN44006) are distributed outside the six faces of the DCT, to measure temperatures with a resolution of approximately 0.5$^{\circ}$C. 


\subsection{Design Performance}
For each triggered event, multiple RoIs can be generated at each wire end.  These RoIs are packaged into data objects that include timestamps, integral charge data, and sampled waveforms.  The drift-time measurement for a given wire can be computed from wire-end timestamps or by applying more sophisticated pulse-shape fitting.\\

The 90:10 CO$_2$:Ar mixture used in flight has a drift velocity of approximately 6 mm/$\mu$s at 20$^{\circ}$C and 14.6 psia in the 0.73 kV/cm field. The diffusion coefficient for this gas mixture is low, very near the thermal limit due to the predominance of CO$_2$. This results in an expected diffusion-limited resolution of better than 70 $\mu$m per point for tracks from particles with  $Z>3$. \\

In the non-bending plane, the particle-crossing position along the wire can be calculated by comparing the charges from opposite wire ends using the charge-division method. A single-hit resolution of 5 mm has been obtained from magnet-off $Z=2$ data collected during the flight. This agrees well with expectations.\\

\section{Time-of-flight System}
\label{sec:TOF}

The HELIX instrument relies on a system of counters made from sheets of plastic scintillator to provide its trigger and to measure the velocity and charge of each particle passing through it.\\

\subsection{Scintillators}
The system has three layers. 
The top and bottom layers each measure 1.6~m x 1.6~m and are located above the magnet and below the RICH, as shown in Fig.~\ref{instrument}.
They are separated vertically by 2.3 m, the lever arm over which the velocity measurements are made. 
The middle layer (`bore paddle') measures 0.61~m x 0.61~m  and is located 1.5 m below the top layer, at the lower end of the magnet bore,  just above the
radiator of the RICH.
It provides a trigger-level requirement that particles pass through the drift-chamber tracker.\\

All counters are made from 10-mm-thick EJ-200 plastic scintillator (Eljen Technology, refractive index $n=1.58$, attenuation length $\lambda = 380$~cm).  
The material is based on polyvinyltoluene and has an emission spectrum that peaks at 425 nm. 
Light-pulse rise time is 0.9 ns and decay time is 2.1 ns.\\

\begin{figure}[htbp]
\centering
\includegraphics[width=0.9\textwidth,angle=0.]{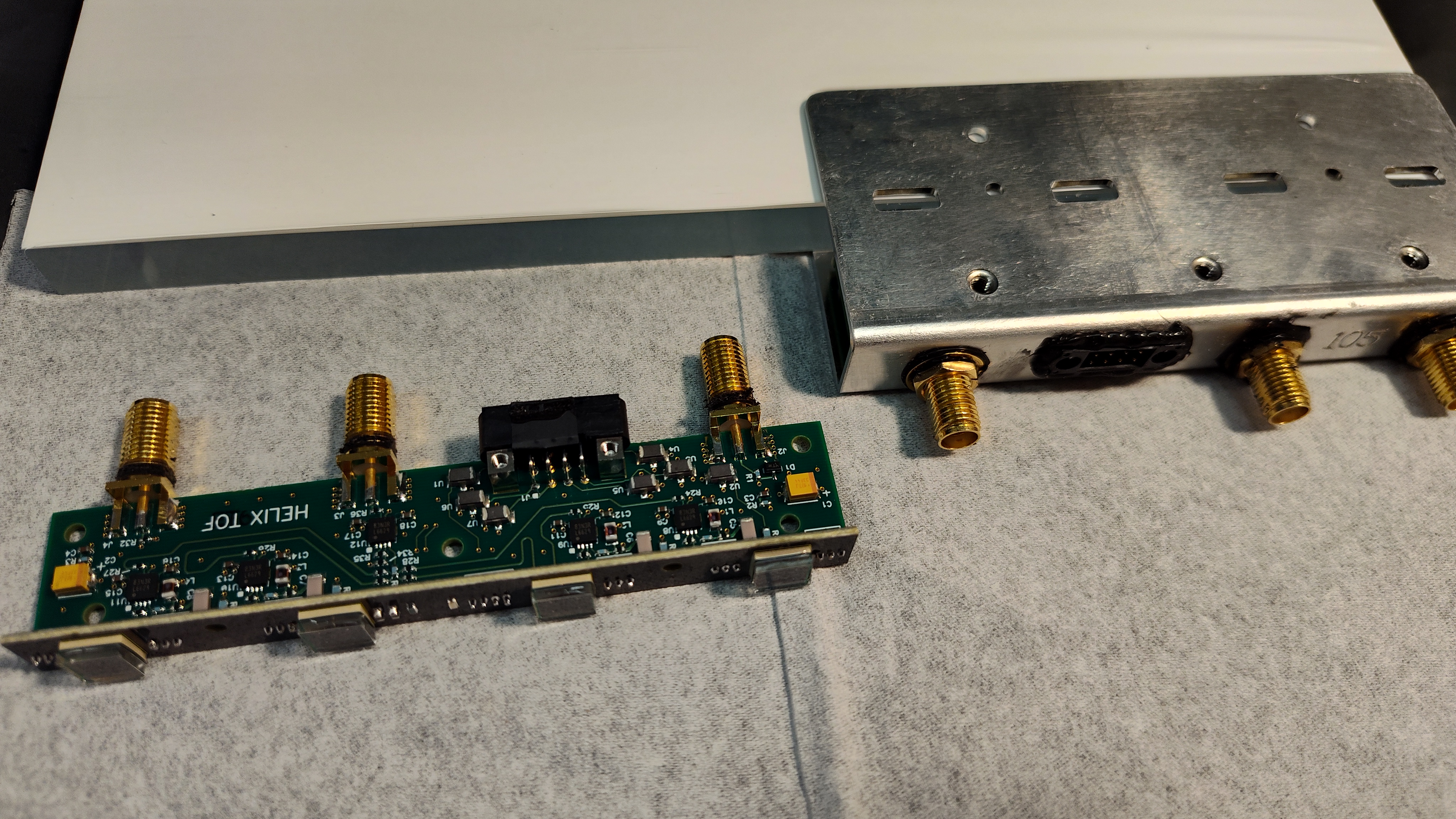}
\vskip 0cm
\caption{
Close-up view of a
time-of-flight scintillation counter end, with two groups of four SiPMs.
At left, one sees four SiPMs, with silicone-rubber cookies attached, mounted on a vertical printed-circuit board which is attached to an electronics front-end board. 
This assembly is inserted into a metal casing and placed against the scintillator's edge, as can be seen on the right.
Inputs and outputs include three SMA connectors for fast signal out, slow signal out and LED pulse in, and an 8-pin miniature connector (Harwin Datamate) for power, temperature readback, and bias voltage.
The scintillator shown here is wrapped in Teflon, but not yet in Tedlar.  
}
\label{toffees}
\end{figure}

The top and bottom layers each comprise eight identical counters, 0.2 m wide by 1.6 m long, arranged side by side. 
Light from each of the eight counters is recorded by 16 Hamamatsu S13360-6050VE silicon
photomultipliers (SiPMs), eight at each end, as can be seen in Fig.~\ref{toffees}.
The middle layer is made from a single sheet, instrumented with 16 SiPMs on each of two facing edges.
The SiPM assemblies are mechanically attached to the scintillator counters using two-component silicone rubber (Momentive RTV615). Optical coupling of the SiPMs to the scintillator material is achieved with cookies made of optical silicone rubber (Eljen Technology EJ-560) and visible in Fig.~\ref{toffees}.\\


All counters are individually wrapped in three layers of extruded white Teflon (polytetrafluoroethylene) film (75 $\mu$m thick), to
improve collection of the scintillation light not captured by total internal reflection, and then further
wrapped in three layers of black DuPont Tedlar (polyvinyl fluoride) film (47~$\mu$m thick) for light-tightness.\\

\subsection{Silicon Photomultipliers and Electronics}

The SiPMs each have an active surface 6~mm x 6~mm, with 50-$\mu$m microcells, and are, by their nature, immune to residual fields from the HELIX magnet.
They feature a large dynamic range and high photo-detection efficiency (40\% at 450 nm), a fast (0.9 ns) rise time, and low transit-time spread. 
Operating voltages are less than 60 V, so there are no breakdown problems resulting from the low atmospheric pressures encountered during flight.\\

\begin{figure}[htbp]
\centering
\includegraphics[width=0.7\textwidth,angle=0.]{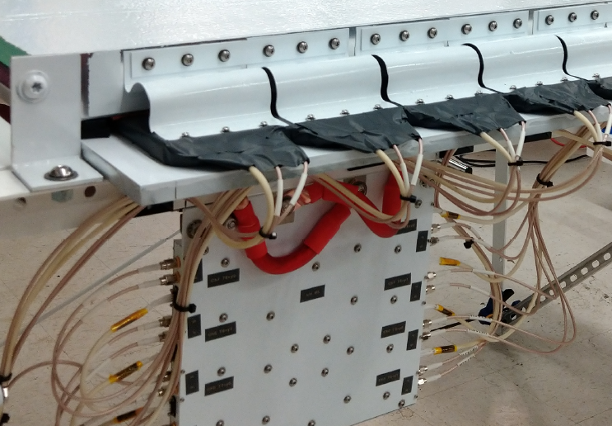}
\vskip 0cm
\caption{
A photograph of two ToF-counter ends, each with two groupings of four SiPMs, cabled to a single readout module mounted vertically below. The metal casing of each SiPM assembly shown in Fig.~\ref{toffees} is thermally coupled to the instrument frame using the white aluminum strips attached to a horizontal bar. The curvature of the soft metal strips allows for possible motion of the scintillator ends due to thermal expansion or contraction. The cabled SiPM assemblies are wrapped in black Tedlar to keep the counters light-tight.
}
\label{strips}
\end{figure}

Like all SiPMs, they are noisy, with dark-current rates of order 2 MHz per channel at room temperature, so temperature control is important to permit operation at low thresholds.
The SiPMs are mounted on, and thermally coupled to, front-end readout boards (see Fig.~\ref{toffees}) which are themselves thermally coupled to the instrument frame using aluminum 1100 strips, as shown in Fig.~\ref{strips}. With this scheme, no active cooling of the SiPMs is necessary.
Despite the SiPM noise rate, the coincidence condition, requiring at least one hit from each
of the top, bore and bottom scintillator layers, results in a very clean event set, with muons at sea level
being recorded at 25 Hz, with over 99\% efficiency. \\

\begin{figure}[htbp]
\centering
\includegraphics[width=0.7\textwidth,angle=0.]{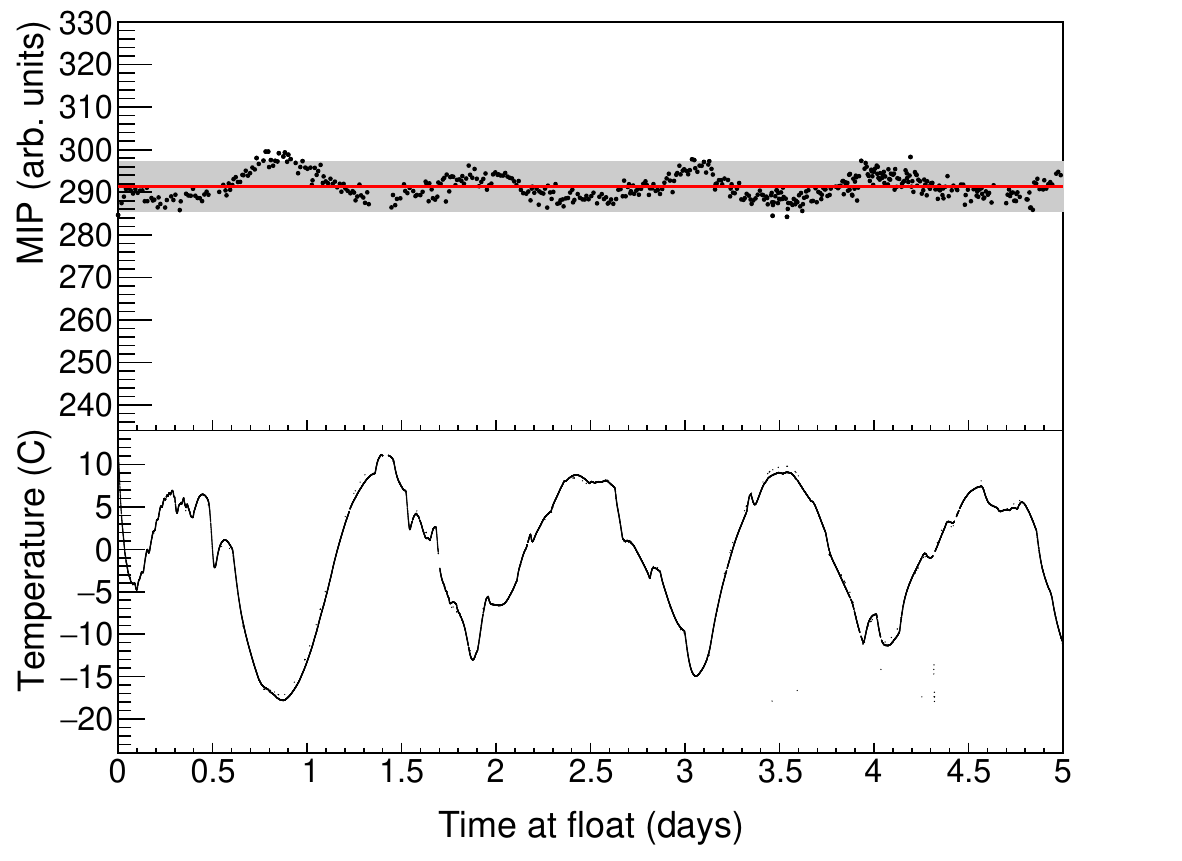}
\vskip 0cm
\caption{
SiPM gain (top panel) and temperature (bottom panel) as a function of time at float for a 
typical SiPM group on a top ToF counter. The proxy for SiPM gain is the most probable ionization energy deposited in the scintillator by a cosmic-ray proton (a minimum-ionizing particle, or MIP) in arbitrary ADC units. The average value is indicated by the solid line and a $\pm 2\%$ variation is shown by the gray band. While some SiPM groups experienced wide temperature excursions in flight, their gains remained stable.
}
\label{gaintime}
\end{figure}

SiPMs are connected in groups of four to front-end readout boards (as seen in Fig.~\ref{strips}). Each board has a temperature sensor and circuitry that can dynamically alter bias voltages to maintain a constant gain (averaged over the four SiPMs). This is illustrated in Fig.~\ref{gaintime} which shows the gain and temperature as a function of time at float for a typical top detector SiPM assembly (where the temperature variations are largest, here ranging over nearly 30$^\circ$C). The SiPM gain remained stable at the level of $\pm 2\%$ throughout the flight, with a small residual variation that is weakly anticorrelated with temperature, and which is tracked and corrected for.\\

The four SiPM signals are passively summed and are used to generate a slow signal for charge measurements and triggering, and a fast signal for timing purposes. The faintest signals occur for a singly-charged particle (e.g., a proton) impacting near the end of a counter, leading to at least tens of detected photoelectrons by a SiPM group at the opposite end (160 cm away). Trigger thresholds are adjusted to the level of a few photoelectrons, well below the faintest signal.\\

The timing signal is connected to a comparator which drives the clock input of a fast flip-flop, the output of which is connected to a time-to-amplitude converter sampled by 
a 14-bit FADC running at 
40 MSPS, as well as a secondary comparator used to toggle the flip-flop data input to allow for high-rate operation. The comparator
has a programmable hysteresis level to ensure the threshold
crossing is reliably captured by the flip-flop, and the secondary comparator toggles the flip-flop data input after a programmable time delay. When the flip-flop toggles, a timing ramp is generated, from either a low to a high state (an `up' ramp) or the reverse, a high to a low state (a `down' ramp), and it is this ramp signal that is sampled at 40 MSPS. Allowing the ramp state to hold after an event is digitized reduces the dead time that would be induced by waiting for the ramp to be reset after each event.
The net effect is that consecutive events have alternate ramp polarities. \\

\begin{figure}[htbp]
\centering
\includegraphics[width=0.9\textwidth,angle=0.]{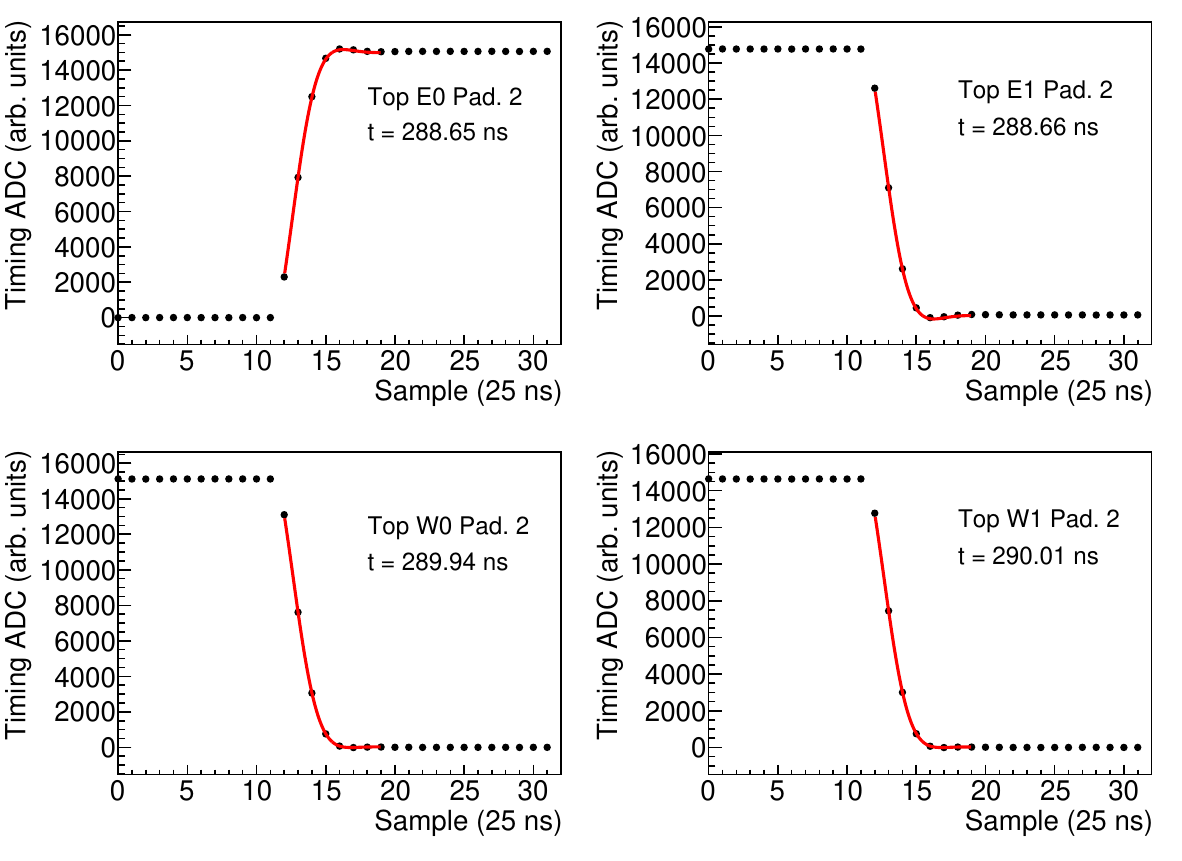}
\vskip 0cm
\caption{
  Timing ramps generated and sampled at 40 MSPS in each of the four SiPM FEE assemblies in a top counter from the transit of a candidate oxygen nucleus recorded in flight. The top panels display the timing signals in the FEE assemblies on the East side of the instrument (E0 and E1), and the bottom panels display the West signals W0 and W1 at the opposite end. Each point represents a 25 ns sample of the waveform, and whether an `up' ramp is generated (E0) or a `down' ramp (E1, W0, W1) depends on the state of the circuit at the time of the trigger. Each ramp is fitted with a phenomenological function (curves in each panel).
}
\label{timingramps}
\end{figure}

Examples of timing ramps generated in a top scintillation counter by the transit of a candidate oxygen nucleus are shown in Fig.~\ref{timingramps}. The digitized waveform samples (each 25 ns long) are displayed as data points, for the two East-end SiPM FEE assemblies (E0 and E1, top panels) and the two West-end FEE assemblies (W0 and W1, bottom panels). Whether an `up' or a `down' ramp is generated depends on the state of the ramp generating circuit at the time of the trigger. Phenomenological function shapes are derived from circuit simulations and adjusted for each readout electronics channel. They are used to fit the measured ramps  with a single degree of freedom, a horizontal timing shift representing the time measurement by each FEE assembly (see the fitted curves in Fig.~\ref{timingramps}). The timing values so derived are indicated on the figure. The values being near 289 ns is not meaningful, as they refer to the main instrument trigger time. The fact that the two signals from a given end agree with each other within a fraction of a nanosecond is important. In this example, the East signals being earlier than the West signals indicates that the cosmic-ray nucleus transited the counter closer to its East end.
The time-to-digital resolution achieved is approximately 25 ps, ensuring that the overall ToF time resolution is driven by the scintillator performance.
For singly-charged particles (e.g., ground-level muons), the top-to-bottom transit timing resolution achieved is approximately 200 ps at present, an evolving and improving figure as the data analysis matures. For Be isotopes, which have four units of charge and therefore generate more light, this is expected to reach a level  better than the design target of 50 ps.\\


\begin{figure}[htbp]
\centering
\includegraphics[width=0.9\textwidth,angle=0.]{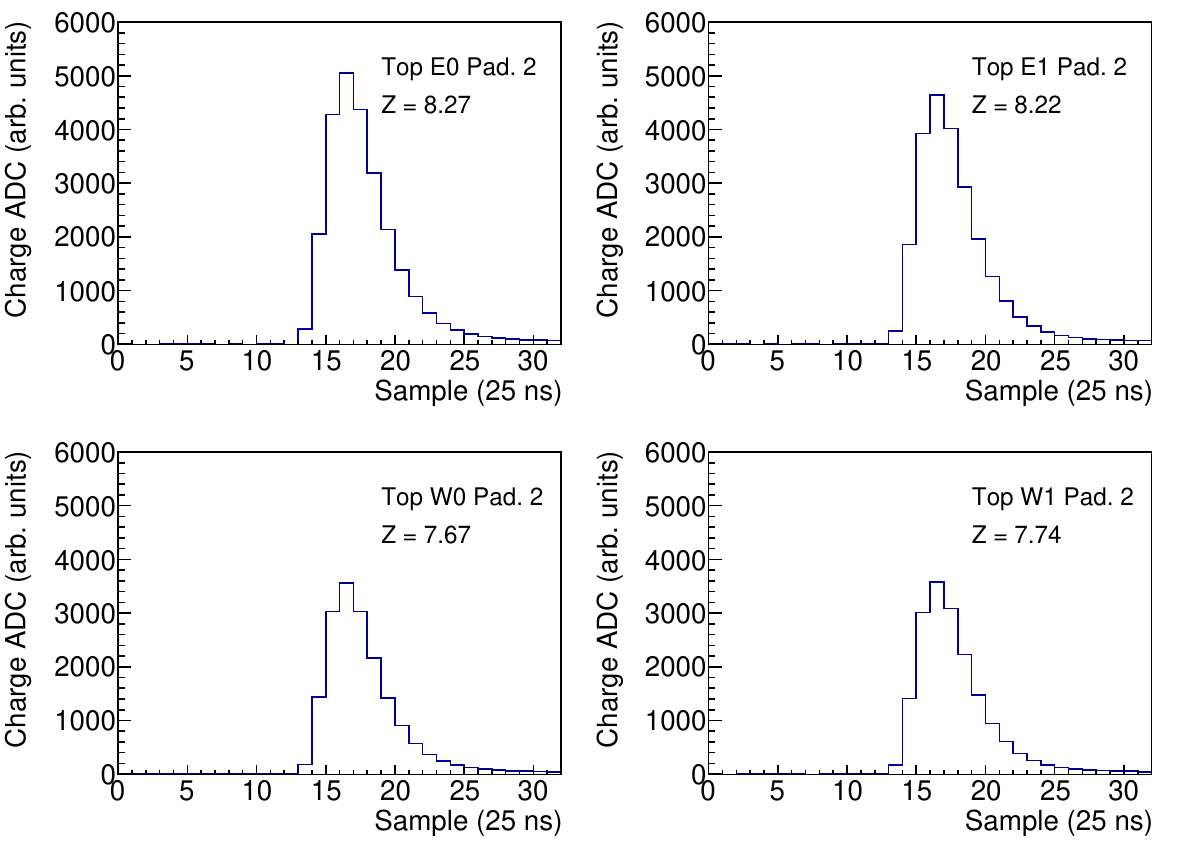}
\vskip 0cm
\caption{
Slow charge signals generated and sampled at 40 MSPS in each of the four SiPM FEE assemblies in a top counter from the transit of a candidate oxygen nucleus recorded in flight, the same event used in Fig.~\ref{timingramps}. The top panels display the charge signals in the FEE assemblies on the East side of the instrument (E0 and E1), and the bottom panels display the West signals W0 and W1 at the opposite end. Each histogram bin represents a 25 ns sample of the waveform.
}
\label{chargepulses}
\end{figure}

The slow signal from each FEE is connected to a 14-bit FADC, of the same type as used for digitizing the timing ramps and running at 40 MSPS, as well as to a pair of comparators driving flip-flops used for triggering (see Sec. \ref{sec:Trigger}). Thus, a waveform is recorded instead of a single integrated charge estimate.
The digitized slow signal, a measure of the light level produced in the scintillator, is used for estimating the charge of the transiting particle.
This is illustrated in Fig.~\ref{chargepulses} which shows the digitized pulses from the same four FEE assemblies and the same event as in Fig.~\ref{timingramps}. Each panel indicates the integrated pulse signal normalized to a $Z=1$ pulse signal, thus indicating the nucleus charge number, here close to 8 as expected for an oxygen nucleus. That the East signals are larger than the West signals indicates that the transit of the cosmic ray occurred closer to the East end of the counter, in agreement with the effect in Fig.~\ref{timingramps} where the East signals are earlier than the West ones. 
The system is designed to detect and identify the charge of nuclei from protons ($Z=1$) to neon ($Z=10$) with a resolution of $0.2~e$.\\


Both the timing and triggering portions run independently, with timing thresholds set low to minimize the effects of time-walk error. The single-channel trigger rates at float altitudes are typically on the order of $\sim 50-100~\mathrm{kHz}$. The FADC outputs are digitally delayed by up to $800~\mathrm{ns}$ to align the data. The FADC data are then stored, to be forwarded to the data acquisition system upon receipt of the global experiment trigger. \\


\section{Ring-imaging Cherenkov Detector}
\label{sec:RICH}

At energies above approximately 1~GeV/n, the velocities of cosmic-ray nuclei are too close to $c$ for the ToF system to be of use in determining their mass. It is here that the HELIX ring-imaging Cherenkov (RICH) detector is used.
Particles traverse a planar radiator, giving rise to the emission of Cherenkov light in the form of a cone with a velocity-dependent opening angle.
Photons in this cone impact an array of SiPMs 500~mm below the radiator, forming a conic section that can be fitted in order to extract the velocity.\\

\noindent
\subsection{Radiator}

\begin{figure}[htbp]
\centering
\includegraphics[width=0.9\textwidth]{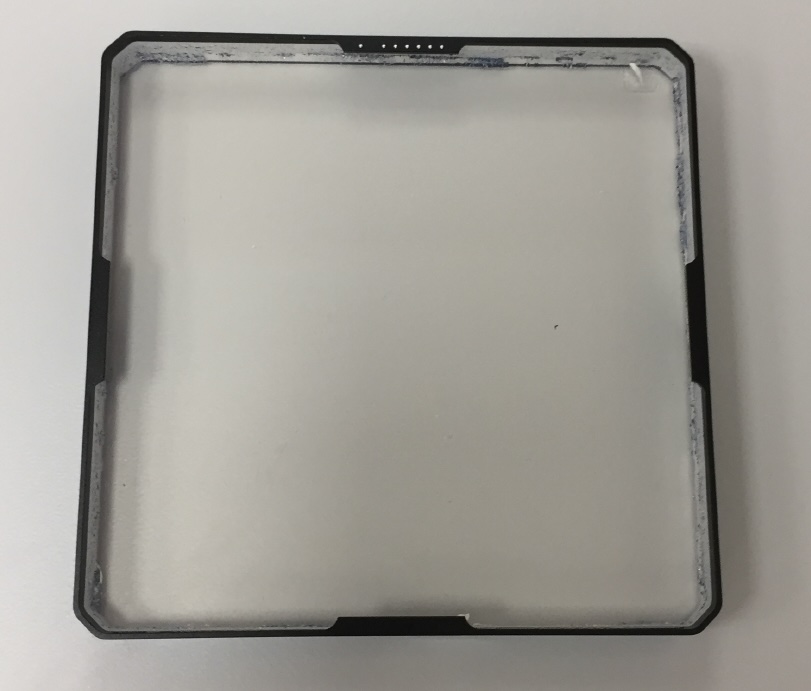}
\vskip 0cm
\caption{An aerogel tile mounted in its aluminum frame. The walls of the 
frame are 1~mm thick and the four visible tabs ensure that the tile cannot 
become loose during flight. 
}
\label{tile}
\end{figure}

\begin{figure}[htbp]
\centering
\includegraphics[width=0.9\textwidth,angle=270.]{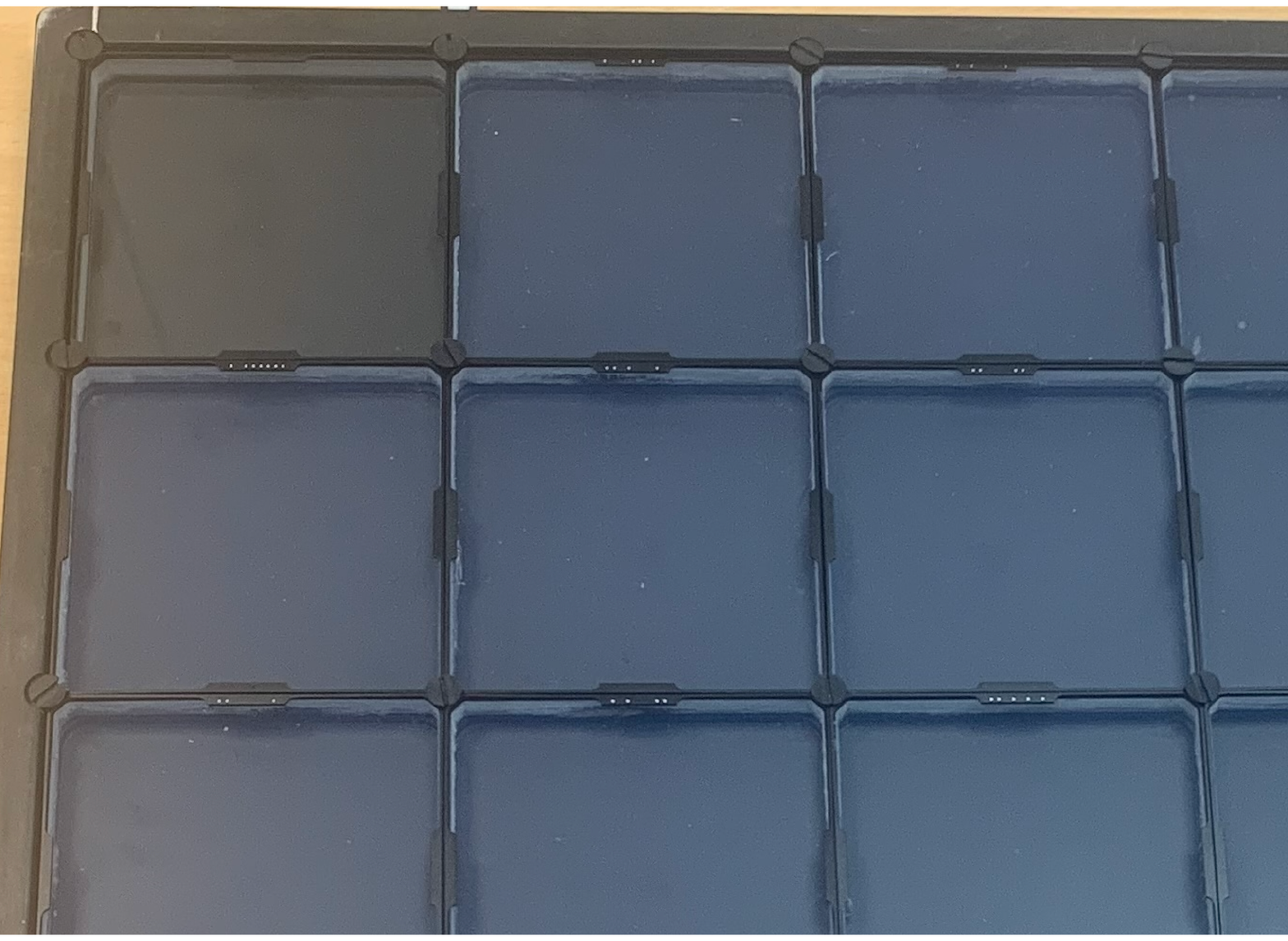}
\vskip 0cm
\caption{A view of one corner of the RICH radiator array. All tiles are secured to the support plate with nylon screws. The corner tile (top-left in the photo) is a NaF crystal and the others are made from silica aerogel. The dots on the tile frames are for identification.
}
\label{tray-corner}
\end{figure}

The radiator consists of a $6 \times 6$ array of tiles, each nominally 100~mm $\times$ 100~mm $\times$ 10~mm in size.  The four corner tiles are polished NaF crystals (Korth Kristalle GmbH), with refractive index 1.33, which are used for cross-calibrating the RICH with the ToF system. The remaining tiles are made from hydrophobic silica aerogel and have a refractive index of approximately 1.15. These were made at Chiba University using a novel pin-hole drying technique
\cite{2020NIMPA.95261879T, 2017arXiv170100143T}. As shown in Fig.~\ref{tile} and Fig.~\ref{tray-corner}, the tiles are housed in aluminum frames and are secured to an aluminum tray, using nylon flat-head screws, for deployment in the HELIX detector. The frames, 1~mm thick and 12~mm deep, provide hard edges for purposes of metrology and to protect the aerogel edges.  Small tabs prevent the tiles from coming loose once mounted on the radiator tray.\\

\begin{figure}[htbp]
\centering
\includegraphics[width=0.99\textwidth]{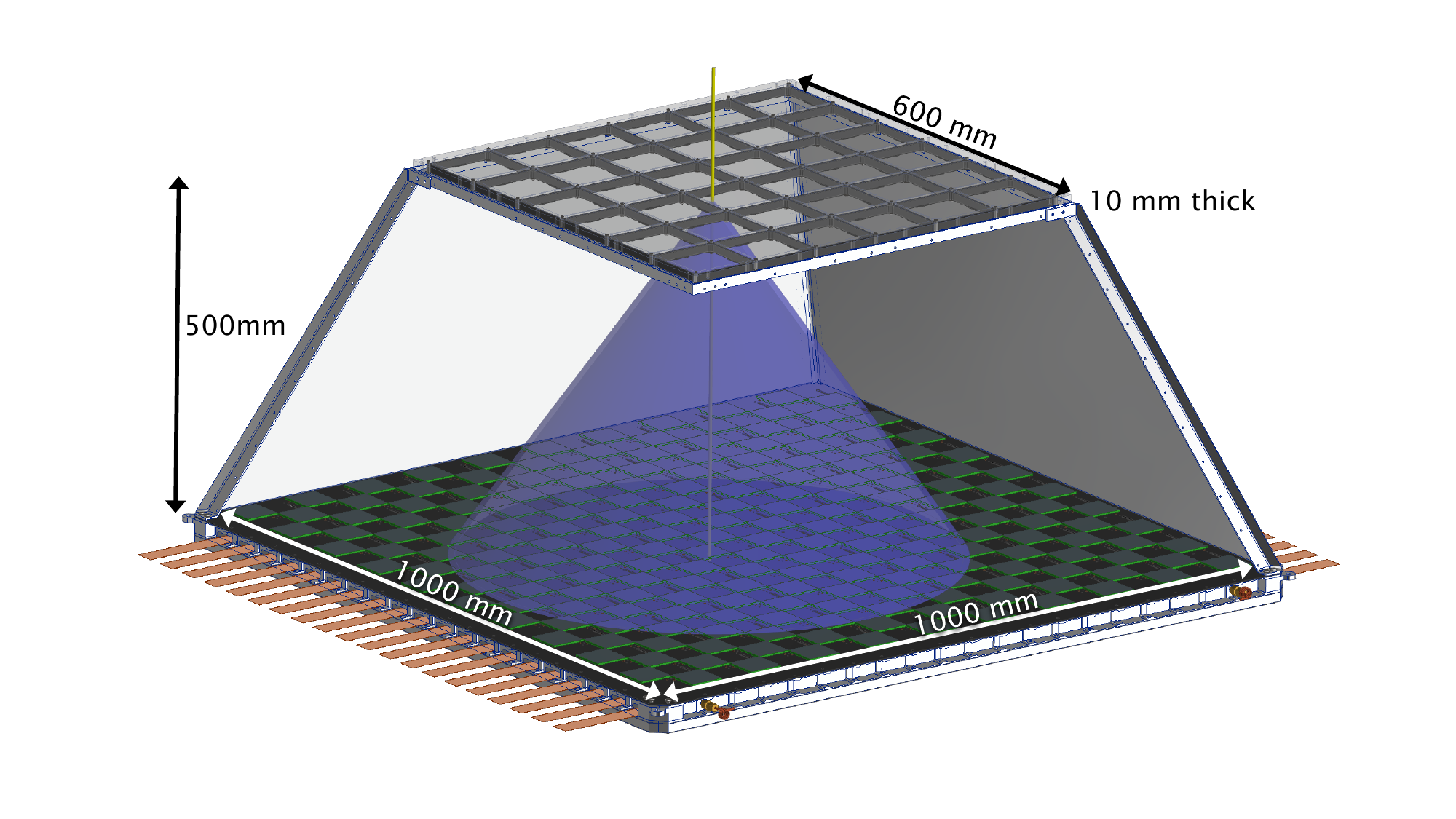}
\caption{A rendering of the HELIX RICH. The radiator plane is separated by 500~mm from the focal plane by a light-tight truncated pyramid constructed from 20-gauge ($\sim$0.8 mm) aluminum sheet, painted black on the interior.
}
\label{rich-assembly}
\end{figure}

\noindent
\subsection{Focal Plane}

The radiator and focal plane are separated by 500~mm and enclosed by a light-tight truncated pyramid, as shown in Fig.~\ref{rich-assembly}. This assembly is mounted under the bore counter, immediately below the magnet cryostat.\\

The focal plane is instrumented with 200 Hamamatsu S14498 SiPM arrays, custom-designed for HELIX and laid out in a checkerboard pattern of 50\% occupancy, for budgetary reasons. 
Each array consists of 64 pixels in an $8\times8$ configuration.
The pixels are each 6~mm $\times$ 6~mm and have 75~$\mu$m microcells. Their photon detection efficiency is approximately 55\% at 450 nm.\\

The SiPM signals are routed to front-end electronics boards (FEEs) mounted at the edge of the payload 
gondola using 700~mm-long flexible PCB cables fitted with Hirose FX11B-100S-SV connectors. 
Each front-end board contains 16~CITIROC~1A application-specific integrated circuits (ASICs)~\cite{CITIROC}, each with 32 channels, allowing 8~SiPM arrays (512 channels) to be processed.
Within the ASIC, the output of each SiPM pixel is amplified and split into fast-trigger and charge-output paths. The trigger path is used to collect timing information on photon hits with a resolution better than 12.5~ns; this can be used to suppress dark-count hits. The charge path feeds a slow shaper followed by a sample-and-hold circuit which provides multiplexed pulse-height outputs to an external 12-bit~ADC following an instrument trigger.  ASIC interfacing, data packaging, and communications are handled by a pair of Xilinx Artix-7 FPGAs, effectively splitting the physical board into two nearly independent logical boards.  \\

\begin{figure}[htbp]
\centering
\includegraphics[width=0.9\textwidth]{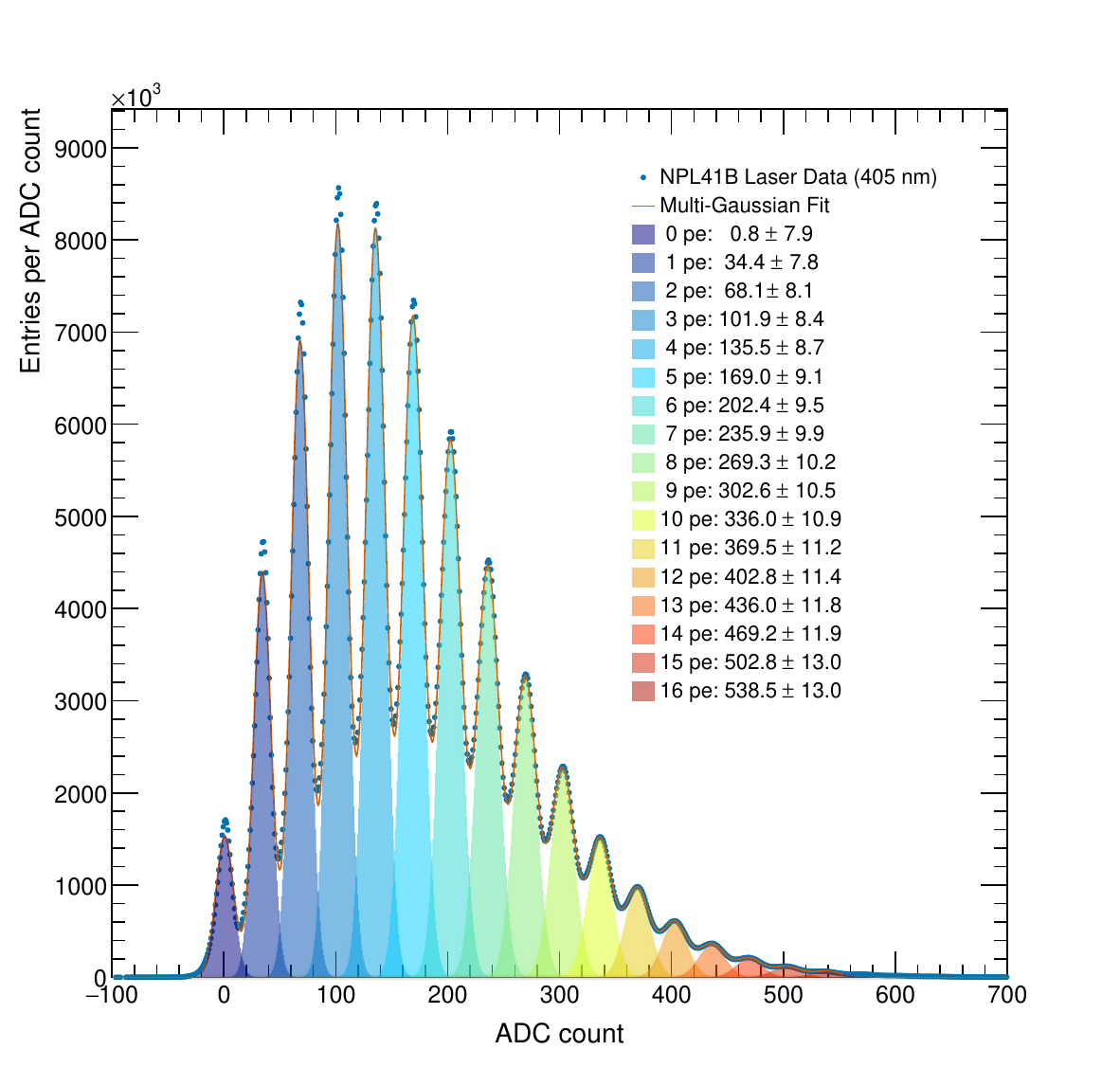}
\vskip 0cm
\caption{A superposition of photoelectron signals of the gain-matched SiPM channels of the RICH focal plane with multiple Gaussian fits representing various pe counts.
}
\label{SiPM-gain-match}
\end{figure}

The bias voltage for the SiPMs (nominally 41~V) is provided by two filtered RECOM R05-100B isolated DC/DC converters. The voltages are bussed to the front-end boards and distributed to the SiPMs using the flex cables described above. Using internal CITIROC trim circuits, the bias for each SiPM array can be adjusted on a channel-by-channel basis.  The voltages are adjusted to provide uniform gains, as determined by special calibration runs wherein the SiPMs are illuminated by light flashes from a diode laser (NPL41B, Thorlabs) delivered through optical fibers terminated in cannula diffuser tips (CFDSB05, Thorlabs). There are four such diffuser tips installed, one on each side of the light-tight pyramid. Fig.~\ref{SiPM-gain-match} shows a plot of pulse-heights summed over the entire SiPM array, illustrating the quality of gain-matching achieved.\\

Each SiPM array board has two Texas Instruments LMT70 sensors which are used to monitor temperatures during flight.  Prior to each data run (of typically 100,000 to 300,000 events over no more than 15 minutes), temperatures are assessed and bias voltages adjusted to maintain gain uniformity through the flight.\\


\begin{figure}[htbp]
\centering
\includegraphics[width=1.0\textwidth]{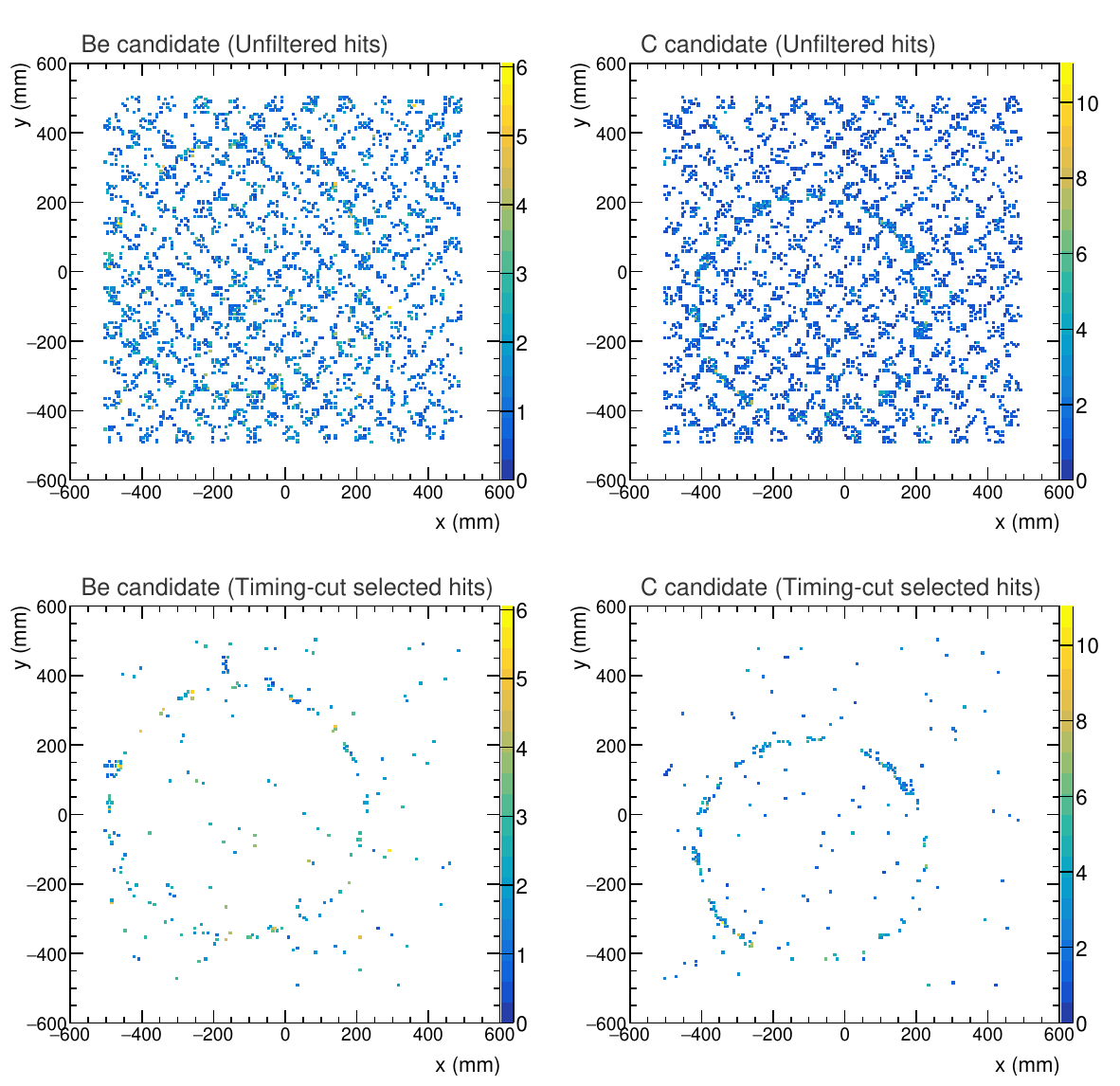}
\vskip 0cm
\caption{RICH SiPM signals (in photoelectrons) before and after a timing cut designed to suppress thermal noise encountered during the 2024 engineering flight. The identification of candidate chemical species is performed using charge information from the ToF. \\
}

\label{rich-rings}
\end{figure}

The RICH readout boards can trigger with a threshold as low as 0.3 photo-electrons (pe) and can provide pulse-height information over the range from 1 to 100 pe.
To reduce dark current, the temperature of the focal plane is regulated using thermoelectric coolers (TECs) mounted on the edges of the focal plane. The hot side of the TECs are cooled using a coolant loop (containing Dynalene HC heat-transfer fluid) that connects to a passive radiator system made from a thick aluminum plate mounted outside the payload gondola. The cooling system was not sufficient to overcome the heating effects encountered during the 2024 engineering flight, resulting in a higher than expected dark rate and hence, signal occupancy. Nonetheless, a timing cut can be employed to suppress background hits, making it easier to distinguish Cherenkov rings. Some examples of this can be seen in Fig.~\ref{rich-rings}. \\

\noindent
\subsection{Calibration}

Manufacturing tolerances inherent in the aerogel production process result in spatial variations of the index of refraction within each tile. These, along with non-planar surfaces that impact the expansion distance for the Cherenkov cones, need to be precisely measured in order to achieve the design goal of 2.5\% mass resolution at 3 GeV/n, sufficient to resolve the beryllium isotopes. 
A precise knowledge of the refractive index, at the level of $\Delta n / n \simeq 7 \times 10^{-4}$ is required.\\

The surfaces of all the aerogel tiles were measured using a Mitutoyo QC606 coordinate measuring machine, and their refractive indices were measured using a 35 MeV electron beam. The tiles all exhibit bowing; the top and bottom faces are roughly parallel but curve slightly in a way that is well described by a two-dimensional parabola. Bowing amplitudes of order 1 mm are not uncommon. Deviations in the index of refraction of order $10^{-4}$ over the extent of a given tile were found but are smoothly varying and can also be parametrized by a two-dimensional parabola.
For details, see reference~\cite{2023NIMPA105568549A}.\\

\noindent
\section{Hodoscope}
\label{sec:hodo}




To achieve the best performance from the RICH, it is desirable to measure the trajectory of the incoming particle as well as possible. This not only improves the reconstruction of Cherenkov angles from focal-plane hits, but allows for increased precision in calculating radiator index and curvature information from calibration data. While the DCT provides excellent reconstruction in the yz (bending) plane, it relies on less-precise charge-division methods in the xz (non-bending) plane. To provide an extra constraint in this projection, a hodoscope detector based on scintillating optical fibers has been developed and is deployed immediately above the RICH radiator plane.

\noindent
\subsection{Design}

\begin{figure}[htbp]
\centering
\includegraphics[width=0.6\textwidth]{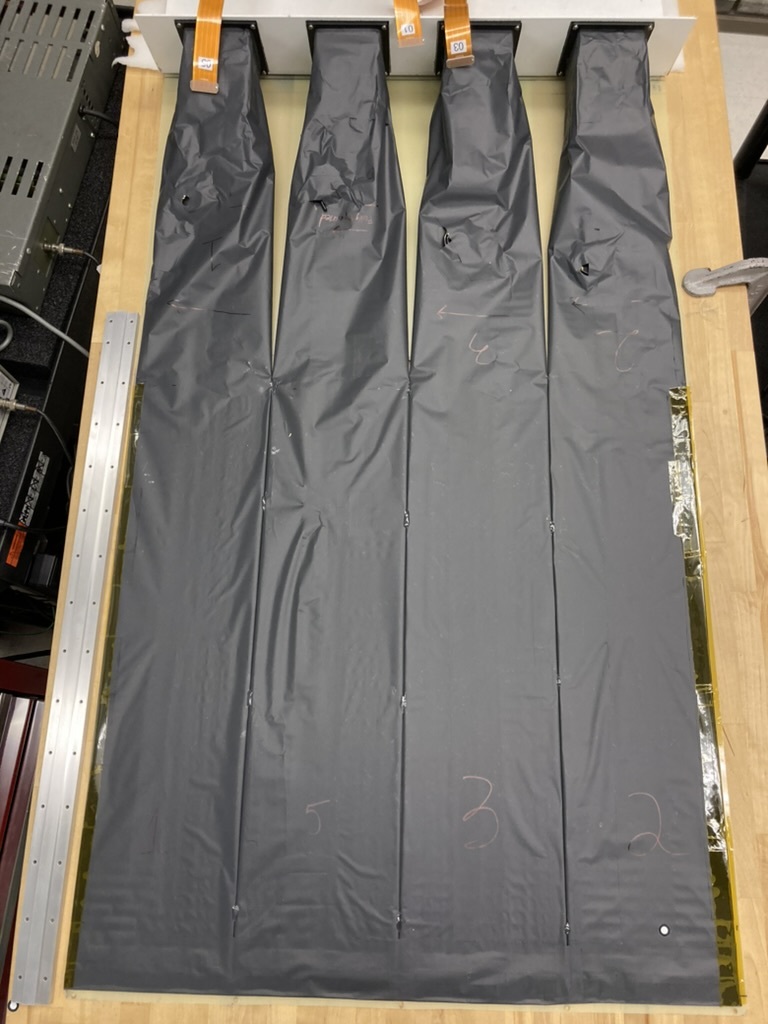}
\vskip 0cm
\caption{
  The fiber hodoscope before installation into the HELIX payload. Four ribbons, each with 150 scintillating fibers, cover the 600~mm $\times$ 600~mm area of the RICH radiator. The fibers are 1000~mm in length, with 400~mm serving as light guides, connecting to four SiPM arrays for readout. 
}
\label{ribbons}
\end{figure}

The hodoscope comprises four ribbons of contiguous optical fibers made from plastic scintillator material (BCF12 from Saint-Gobain Crystals - now Luxium Solutions) and made light-tight with Tedlar sheets, as seen in Fig.~\ref{ribbons}.
Each fiber is square in cross-section, 1 mm on a side, with a core of polystyrene (refractive index $n=1.60$) and a cladding of acrylic ($n=1.49$), approximately 40 $\mu m$ thick.  Each ribbon has 150 fibers, making the total width of the hodoscope 600 mm, to match the width of the RICH radiator plane.
The fibers are 1000 mm in length, although only 600 mm are meant for use in detecting particles passing through the RICH acceptance. The remaining 400 mm length of each fiber is used as a light guide.\\

\noindent
\subsection{Readout}

\begin{figure}[htbp]
\centering
\includegraphics[width=0.99\textwidth]{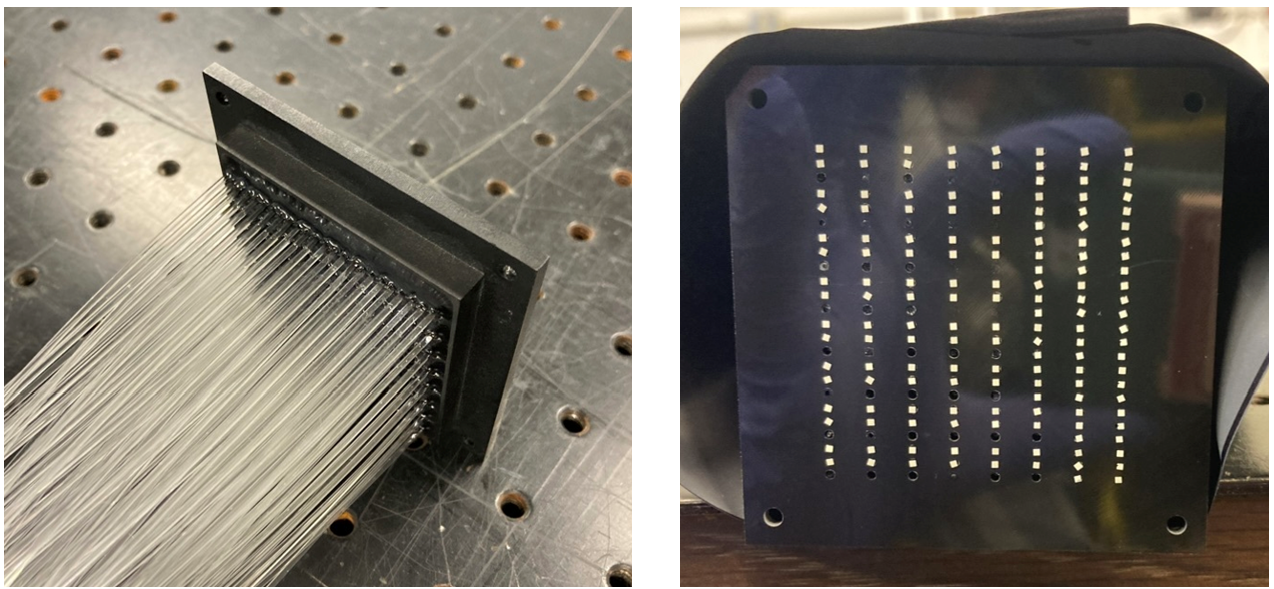}
\vskip 0cm
\caption{
  Illustration of the fiber-to-SiPM coupling. 150 fibers from each ribbon are regrouped such that each of the 64 pixels in an $8 \times 8$ SiPM array sees the light from two or three widely separated fibers.
}
\label{cookie}
\end{figure}

The hodoscope fibers are read out by the same kind of SiPM arrays and associated electronics that are used in the RICH focal plane. As an economy, each of the 64 pixels in a SiPM array is coupled to two or three fibers using a machined Delrin interface device, as shown in Fig.~\ref{cookie}.\\

\begin{figure}[htbp]
\centering
\includegraphics[width=0.99\textwidth]{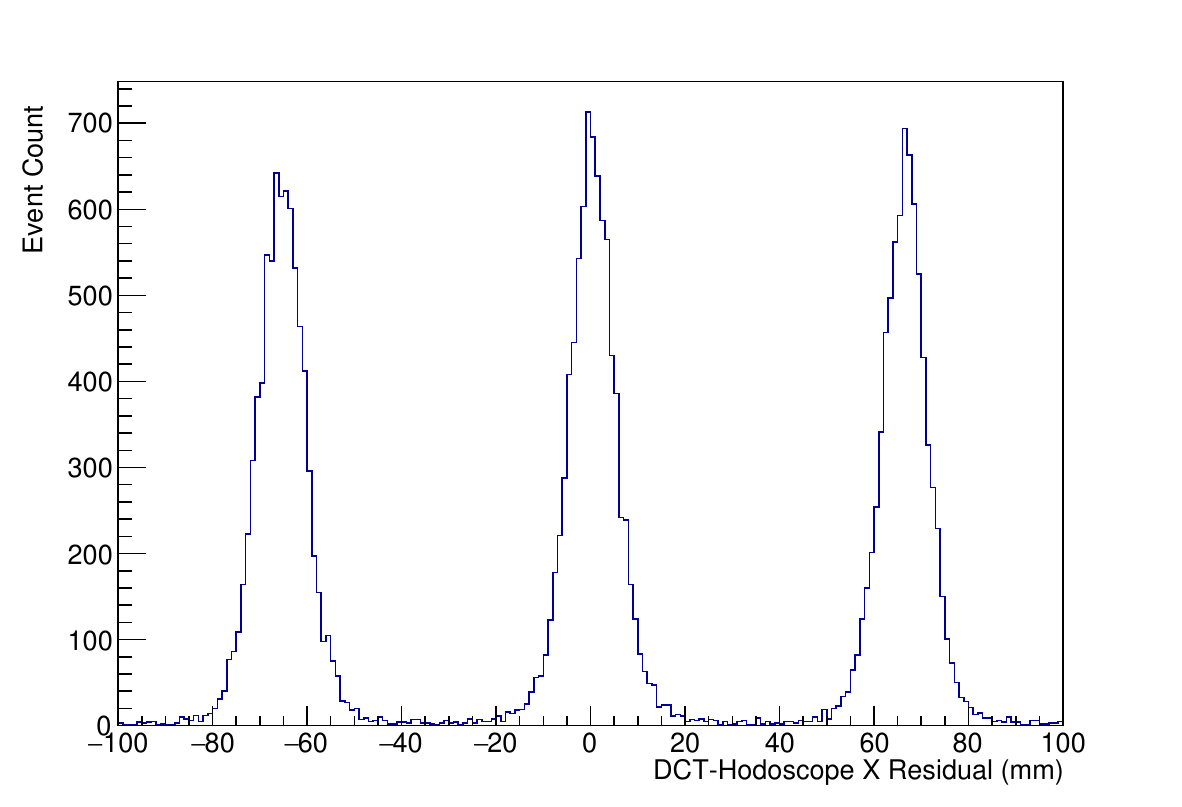}
\vskip 0cm
\caption{Fiber multiplexing performance plot. For events with a single DCT track and a corresponding single SiPM hit in the hodoscope, the x coordinate of the track, extrapolated to the hodoscope plane, is determined. The difference between this value and the x coordinate of the middle of the three fibers coupled to the SiPM is computed and histogrammed. This plot is made with data from three-fiber SiPMs; a similar plot with two peaks can be made with two-fiber SiPMs. The peak widths ($\sigma \sim 5~$mm) result from the precision of the DCT extrapolation and are much smaller than their spacing (64 mm) which is determined by the spacing of fibers routed to a common SiPM.
}
\label{hodoplot}
\end{figure}

Fibers that are coupled to a given pixel are separated from each other 
by 64~mm
so there is little probability of an ambiguous interpretation of the data, given the low multiplicity of the HELIX events. This can be seen in Fig.~\ref{hodoplot}, which was made for events where any SiPM connected to a triplet of fibers has a signal. The distance of where the track intercepted the hodoscope plane, as measured from the position of the central fiber of the triplet is histogrammed. A similar plot for the SiPMs that are coupled only to two fibers can also be made and it exhibits two peaks separated by 64 mm.\\

The peak widths ($\sigma \sim 5~$mm) indicate the precision of the extrapolated DCT tracks in the non-bending plane and provide graphic justification for the hodoscope. Its 1-mm fibers provide a nominal geometric precision of $1/\sqrt{12} \simeq 0.3$ mm. This has been verified using laboratory scans made with a collimated $^{90}$Sr source.

\newpage
\section{Data Acquisition and Trigger\label{sec:Trigger}}

The data acquisition (DAQ) system for HELIX is configured in a distributed star topology.  A central control board termed the `master merger' resides in the payload science-flight computer (SFC) and connects via high-speed links to each of the four primary subsystems: trigger, RICH, ToF, and DCT.  Each non-trigger subsystem has its own merger board that assembles and buffers events from the primary readout boards, described above.  The SFC itself is built around a Perfectron INS8346B mini-ITX industrial motherboard, with an Intel Core i3-3220 chip, and a custom conductive cooling solution.  Flight data were stored on a redundant array of 4~TB Innodisk 3TG-P industrial solid-state disks. \\

\subsection{Data and Commanding}
The primary DAQ interface is the master merger (MM) board, an Ethernet packet analysis engine\footnote{DiNI Group DNPCIe\_40G\_KU\_LL\_2QSFP} containing a Xilinx Kintex-Ultrascale FPGA and two 40~Gbps QSFP+ modules, packaged in a peripheral component interconnect express (PCIe) form factor.  The MM interfaces to each of the four DAQ subsystems via an enhanced small form-factor pluggable (SFP+) interface, using quad SFP (QSFP+) to four SFP+ copper direct-attach links, and a 6.4~Gbps (640~MB/s maximum data throughput) high-speed Aurora serial protocol. The MM firmware is designed to accommodate up to 8 subsystem links for flexibility, with only four currently utilized. \\

The merger boards act as the DAQ interface for their own subsystem. The boards are custom-designed, based on Xilinx Artix-7 FPGAs, with SFP modules to communicate with the MM, as well as multi-gigabit transceivers (MGTs) operating at 1.6~Gbps (160~MB/s throughput) communicating with readout boards. The master trigger board has a similar design, with the same MM interface but without subordinate readout boards.\\

During data-taking operations, the merger boards handle the readout of event fragments from readout boards, assemble those fragments into complete subsystem events, and buffer them. Those subsystem events (of variable length) are transmitted to the MM via the SFP+ link, multiplexed with command and status messages. At the master merger, the subsystem events are demultiplexed from the commands and buffered into the onboard $8~\mathrm{GB}$ random-access memory (RAM) buffer, partitioned to store up to 512 total events with varying sizes for each subsystem. This allows for large subsystem events (up to $2~\mathrm{MB}$ per event), which are used for DCT RoI location tuning during detector setup.\\

Once a subsystem event is available from all enabled detectors, the event and its length are sent separately over PCIe using the Xillybus \cite{xillybus} interface framework to buffer in the SFC system RAM. Events are transferred asynchronously as the internal Xillybus buffers fill. Once sufficient event data are available (typically $\sim 2.5~\mathrm{MB}$ or $150~\mathrm{events}$ during operation), the SFC reads them from the event buffer and saves them to disk. The multi-stage buffering scheme along with the high-speed transfer protocols keep the deadtime contribution from data flows to essentially zero. \\

Command and status packets are also transmitted/received using the Xillybus framework allowing for reading/writing configuration registers and status and monitoring data on the readout, merger, and trigger boards during operation. Each board also allows for firmware updates via this channel.\\

\begin{figure}
    \centering
    \includegraphics[width=0.95\linewidth]{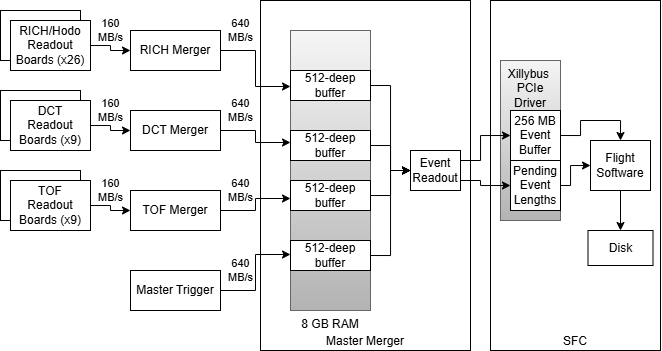}
    \caption{Event data flow for the HELIX DAQ. Data are generated by each
    readout board, collected into a subsystem event at a merger
    (or directly by the trigger), and then sent to the master merger.
    The MM then buffers subsystem events and sends full system events to
    the science-flight computer using the Xillybus framework. Events 
    are read out in bulk by the flight software once sufficient data are
    available.
    }
    \label{fig:daq_data_flow}
\end{figure}

\subsection{Master Trigger Design and Logic}

The master trigger board contains the main clock used by the ToF subsystem, operating at $40~\mathrm{MHz}$ with $0.5~\mathrm{ppm}$ stability over an industrial operating temperature range. This clock is distributed directly 
to the ToF readout boards using a sub-picosecond-jitter clock fanout to maintain timing resolution.
Additionally, each subsystem receives an $80~\mathrm{MHz}$ system clock, derived from the main clock, along with a synchronous system trigger and a reset signal, via the merger boards.
The merger boards also provide subsystem-ready status signals to the trigger to allow for trigger holdoff and deadtime measurements. 
Time is globally synchronized everywhere in the readout electronics using a synchronization
pulse in both the ToF-clock (synchronizing the coarse ToF time) and system-clock (synchronizing event
time) domains.\\

\begin{figure}
    \centering
    \includegraphics[width=0.95\linewidth]{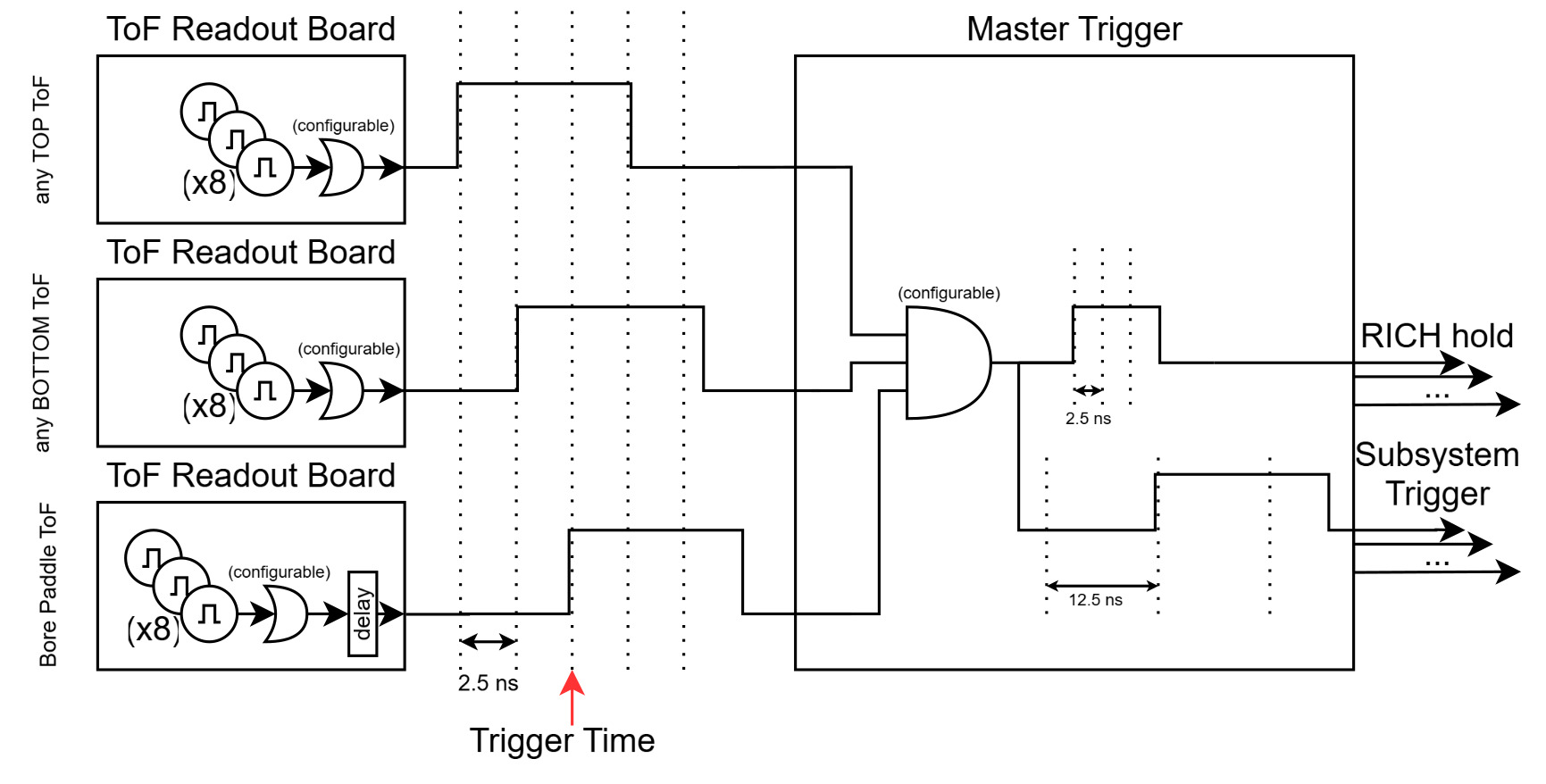}
    \caption{Trigger logic diagram. ToF readout boards generate a trigger pulse derived from their 8 slow inputs (nominally an OR of any input), which are then combined into a global trigger (nominally an AND of any top and bottom counter plus the bore paddle) at the master trigger. The trigger logic runs with $2.5~\textrm{ns}$ quantization to minimize jitter for the RICH analog hold signal, and is resynchronized to the $80~\textrm{MHz}$ system clock for subsystem readout.
    }
    \label{fig:trigger_logic}
\end{figure}

The HELIX trigger (shown in Fig.~\ref{fig:trigger_logic}) is formed by a coincidence between outputs from fast discriminators acting on the `slow' signals at each ToF readout board. Each signal has two variable-threshold discriminators, to generate both large-signal (so-called ZHi) and small-signal (ZLo) triggers. At the ToF readout board, the 8 input channels are combined with configurable logic (nominally an 8-way logical OR) using a time quantization of $2.5\,\mathrm{ns}$.\\

This combined output is then sent to the master trigger as a programmable-length pulse. Signals from the 9 ToF readout boards (4 top, 4 bottom, and 1 for the bore paddle) are then combined using configurable logic to form the dual primary-physics full-instrument ZHi and ZLo triggers. These triggers are combined with additional `auxiliary' (forced, random, and two external-input) triggers to generate a `fast' global trigger (again with $2.5\,\mathrm{ns}$ time quantization). The fast trigger, after a programmable delay, is provided to the RICH subsystem as the `hold' signal for the CITIROC pulse-height output. The fast trigger is also synchronized with the system clock ($80~\mathrm{MHz}$) and sent as the synchronous system trigger to begin the readout process for each subsystem.\\

The configurable trigger logic modules available at the ToF and master trigger allow for different trigger conditions during testing as well as the possibility of ignoring individual channels during flight if necessary. During flight, the logic required a coincidence between a top, bottom, and bore-paddle readout board to form a  trigger, with the coincidence window set by the length of the trigger pulse from the readout boards. The bore-paddle trigger signal was  delayed relative to the top and bottom signals to define the trigger time and eliminate variations due to geometrical effects.\\

The dual trigger scheme allows for the more-abundant $Z=1$ particles to be prescaled at the trigger level in order to reduce data rates.  During the 2024 flight, ZLo triggers were prescaled with a factor between 1.0 (\ie, no prescale) and 4.0 (\ie, about 25\% of $Z=1$ particles accepted).

\subsection{Housekeeping}
An independent data acquisition system was developed for the collection of so-called `housekeeping' (HSK) data, which include temperatures, voltages, currents, pressures, etc.  The hardware for this system consists of five separate  circuit boards, each specialized for the monitoring of a specific subsystem (\eg, DCT or RICH) in the payload, with one board serving as the primary HSK interface to the science-flight computer.\\

The heart of each HSK board is a Texas Instruments TM4C123GXL microcontroller that reads data from local sensors, packetizes and encodes them, and transmits the packets upstream over RS232 for eventual readout by the SFC.  Individual boards not only act as independent devices, but can also act as relays, enabling a daisy-chained network structure.  In normal operations, data packets are sent at fixed intervals to provide a continuous description of the detector and payload state.  The system is also bidirectional, allowing commands originating from the SFC (\eg, temperature-read requests, or relay-toggle requests) to be transmitted to any subsystem in the payload.\\

\subsection{Telemetry}
HELIX employs multiple telemetry channels for data uplink (ground to payload) and downlink (payload to ground).  Below, we describe the configuration used in the 2024 engineering flight.\\

For downlink, the primary channels were high-rate TDRSS (Tracking and Data Relay Satellite System) through a high-gain antenna; line-of-sight (LOS) through a Quasonix EVTM (Ethernet Via Telemetry); and Starlink, through a SpaceX Starlink Maritime antenna.  The line-of-sight channel was active for the first $\usim 12$ hours of the flight and operated at a nominal throughput of $\usim 1$~Mbps during this period.  Starlink was active during the entire flight with nominal throughputs of $4-10$~Mbps.  \\

The actual downlink data consisted of randomly selected events from the data buffers, along with a stream of housekeeping information, with a typical aggregate data rate of $\usim 380$~kbps (as User Datagram Protocol - UDP packets) when the higher bandwidth channels were in use.  These data were collected at one of two active ground-station machines (one at the launch site in Esrange, Sweden and one at CSBF headquarters in Palestine, Texas).  The housekeeping portion of the data was then retransmitted to a cloud database for display in web-based monitoring dashboards.  The event data were collected and distributed to live event displays point-to-point, using the xrootd protocol.  \\

Uplink data consisted largely of payload commands. These were sent from ground-station computers to the payload primarily as UDP packets over Starlink.  Commanding was also possible through multiple routes (\eg, TDRSS, Iridium) provided by the CSBF support instrumentation package. When Starlink connectivity was good, direct ssh access to the payload was  possible. Starlink connectivity continued to function post-landing, however the throughput was not sufficient for significant science data recovery.\\


\begin{figure}[htbp]
\centering
\includegraphics[width=0.8\textwidth]{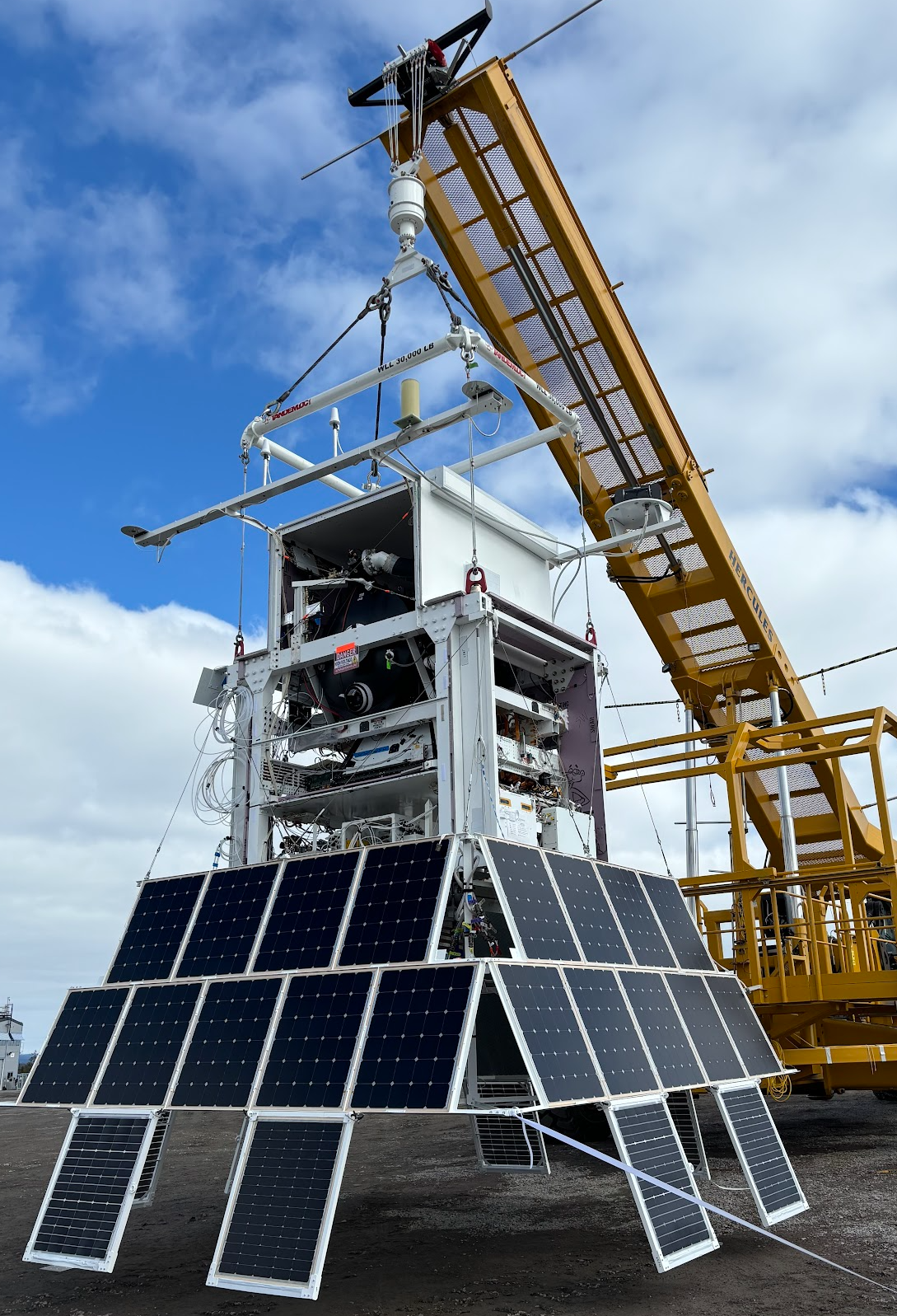}
\vskip 0cm
\caption{
A photograph of the HELIX payload showing the omni-directional skirt comprising four arrays of nine solar panels. A boom for telemetry antennas can also be seen. The photograph was taken prior to launch from the Swedish Space Corporation's Esrange Space Center near Kiruna, before installation of thermal insulation panels on the instrument.
}
\label{payload_hanging}
\end{figure}

\subsection{Power Systems}
\label{sec:power}

The HELIX power system consists of a photovoltaic array (PVA), a set of four charge controllers and a rechargeable battery pack, generating a nominal 24~V system bus voltage.  \\

The photovoltaic array was configured as a 4-sided `omni-directional skirt' hanging below the payload.  This was done as a contingency against a possible loss of pointing ability, due to torques created by the interaction of the HELIX magnetic moment and the Earth's magnetic field.  In the event, the NASA-provided azimuthal rotator was able to overcome these torques, with some adjustments to current limits and pointing feedback parameters.\\

Each side of the skirt consisted of nine solar panels (Suncat Solar, LLC) arranged into three series-connected strings, for a payload total of 36 panels.  
Each of the 0.8~m $\times$ 0.7~m panels
contained 30 SunPower E60 cells connected in series, for a nominal power rating of 100~W at float altitudes.  \\

The three 3-panel strings from each side of the payload, each with a nominal open circuit voltage of approximately 60~V, were fed into a single 1600~W maximum-power point-charge controller (Morningstar TriStar MPPT-60 with thermal modifications) via a parallel Schottky diode OR arrangement.  Both the OR design and the thermal modifications were based on designs by CSBF.  A single charge controller was employed to handle the panels from each side of the payload and they were wired in parallel to the payload battery pack.  That battery pack consisted of two parallel rechargeable lithium-iron-phosphate batteries (Saft Xcelion 6TE), each providing a 26.2~V nominal voltage and an 82~Ah (2100~Wh) nominal capacity, at a mass of 20.7~kg.  \\

Power for the payload was distributed at bus voltage and down-converted as needed at subsystem board locations using ABB (now OmniOn) Hammerhead eighth-brick isolated DC/DC power converters. To reduce the effects of the payload magnetic field, the converters were shielded with 3D-printed shrouds wrapped with 5 layers of 0.1 mm NETIC S3-6 magnetic-shielding foil.    \\

During the 2024 flight, the system performed flawlessly, and was easily able to meet the payload's typical $\sim 870$~W power requirement, under all pointing conditions.  A maximum power-production level load of approximately 1800 W was reached while the batteries were being charged. In addition, the system continued to supply sufficient power to operate the SFC and Starlink for approximately one month after landing, despite the loss of multiple sides of the skirt due to damage from impact or post-landing orientation.

\section{Conclusions and Outlook}
\label{sec:conc}

With its superconducting magnet and high-precision particle tracker providing state-of-the-art rigidity measurements, and a velocity measurement system that relies on time-of-flight at low energies and an aerogel-based RICH at high energies, HELIX is designed to make mass-resolved measurements of light isotopes over a range of energies where different models from cosmic-ray secondary production and propagation feature different behaviors and need to be guided by new data. The HELIX detector, shown in Fig.~\ref{payload_hanging}, was developed over several years and performed very well on a six-day engineering flight in 2024. The data are now being analyzed. The Collaboration plans to implement small improvements, based on lessons learned from the first flight, and undertake a longer flight above Antarctica in the near future.\\



\section*{Acknowledgments}

Primary funding for the HELIX project in the US is provided through National Aeronautics and Space Administration (NASA) grant 80NSSC20K1840. 
In Canada support comes from the Natural Sciences and Engineering
Research Council (NSERC) and the Canadian Space Agency's 
Flights and Fieldwork for the Advancement of Science and Technology (FAST)
program.\\

We thank members of the Columbia Scientific Balloon Facility (CSBF) for their expert contributions during the testing, launching and recovery phases of the 2024 HELIX flight. We are grateful for the warm welcome and technical assistance, provided by the personnel at the Esrange Space Center in northern Sweden, while we prepared the payload for flight.\\ 


Professor Dietrich M\"uller (1936-2021) of the University of Chicago was an early and enthusiastic supporter of HELIX and provided valuable guidance to the Collaboration.

\printbibliography

\end{document}